%% file: ms.tex
\documentclass[12pt,preprint]{aastex}

\newcommand {\thirco}{$^{13}$CO\ }      
\newcommand {\thircons}{$^{13}$CO}

\newcommand {\twelcons}{$^{12}$CO}
\newcommand {\nh}{N$_{2}$H$^{+}$\ }
\newcommand {\nhns}{N$_{2}$H$^{+}$}
\newcommand {\co}{C$^{18}$O\ }
\newcommand {\cons}{C$^{18}$O}

\defcitealias{Kirk06}{KJD06}
\defcitealias{Kirk07}{KJT07}
\defcitealias{Kirk09}{KJB09}
\defcitealias{BB00}{BB00}
\defcitealias{Goodman93}{G93}
\defcitealias{Barranco98}{BG98}

\newcommand{\hknew}{}
\begin{document}

\title{The Dynamics of Dense Cores in the Perseus Molecular Cloud II: 
The Relationship Between Dense Cores and the Cloud}
\author{Helen Kirk\altaffilmark{1,2,3}, Jaime E. Pineda\altaffilmark{4}, 
	Doug Johnstone\altaffilmark{2,1}, and Alyssa Goodman\altaffilmark{4}} 
\altaffiltext{1}{Department of Physics \& Astronomy, University of Victoria, 
        Victoria, BC, V8P 1A1, Canada}
\altaffiltext{2}{National Research Council of Canada, Herzberg Institute of 
        Astrophysics, 5071 West Saanich Road, Victoria, BC, V9E 2E7, 
        Canada}
\altaffiltext{3}{Currently at Harvard-Smithsonian Center for Astrophysics;
	hkirk@cfa.harvard.edu}
\altaffiltext{4}{Harvard-Smithsonian Center for Astrophysics, 60 Garden St,
	Cambridge, MA 02138, USA}


\begin{abstract}
{\hknew We utilize the extensive datasets available for the Perseus molecular
cloud to analyze the relationship between the kinematics of small-scale
dense cores and the larger structures in which they
are embedded.  The kinematic measures presented here can be used
in conjunction with those discussed in our previous work as
strong observational constraints that numerical simulations 
(or analytic models) of star formation should match.  We find that
dense cores have small motions with respect to the \thirco gas,
about one third of the \thirco velocity dispersion along the same
line of sight.  Within each extinction region, the 
core-to-core velocity dispersion is about half of the total
(\thircons) velocity dispersion seen in the region.  Large-scale
velocity gradients account for roughly half of the total velocity
dispersion in each region, similar to what is predicted from
large-scale turbulent modes following a power spectrum of
$P(k) \propto k^{-4}$.
}
\end{abstract}

\section{INTRODUCTION}
Star formation research is entering an era where the properties of dense,
star-forming cores and their environs can be described in a 
statistical manner across many molecular cloud environments.
Existing multi-cloud surveys include the COMPLETE Survey
\citep{Ridge06}, {\it Spitzer} c2d Survey \citep{Evans03}, and 
{\it Spitzer} Gould Belt 
Surveys, while surveys that are in progress 
include the JCMT and {\it Herschel} Gould Belt 
surveys \citep{Wardthomp07,Andre05}.  
These surveys will characterize dense core properties with sufficient
detail to allow for the measurement of both 
global properties as well as the variation
due to molecular cloud environment.  Numerical 
simulations have also begun to make explicit predictions about a 
variety of observable dense core properties, going beyond global statistics 
such as the mass function.

One nearby molecular cloud that has been particularly well-studied
already is Perseus.  The Perseus molecular cloud is located relatively
close-by at a distance of 250~pc \citep{Cernis93} and is an intermediate-mass
star-forming region with protostars and dense cores found primarily
within small clustered environments (e.g., NGC~1333 and IC~348).
To study the dense cores in an unbiassed manner, dust column
density observations have been undertaken in the (sub)millimetre
continuum
(Hatchell et al.~2005; Kirk, Johnstone, \& Di Francesco 2006, hereafter KJD06;
Enoch et al.~2006) and
the large-scale structure of the cloud has been determined
through extinction inferred via reddening of
stars in the 2MASS dataset (KJD06; Ridge et al.~2006). 
The young protostellar content
of the cloud has been measured using {\it Spitzer} data
\citep{Rebull07,Jorgensen07,Hatchell07a}.  Molecular line data also exist on
multiple scales within the cloud.  The kinematics of the
dense cores have been traced by pointed observations of
\nh \citep[][hereafter KJT07]{Kirk07} and NH$_3$ \citep{Rosolowsky08}, and
their immediate less-dense surroundings in \co \citepalias{Kirk07}
and CCS \citep{Rosolowsky08}.  Note that carbon-bearing molecules
tend to freeze out quickly onto dust grains in the centres of cold, 
dense cores, while {\hknew the absence of CO molecules leads to an 
enhanced production of nitrogen-bearing molecules}
\citep[see, for example,][]{Tafalla02}.  Hence in dense cores, nitrogen-based
molecules tend to trace the dense parts of the cores, while 
carbon-bearing molecules tend to trace the lower density
envelopes of the cores.  The less-dense gas in Perseus has been mapped in its
entirety at lower resolution
in \twelcons(1--0) and \thircons(1--0) 
\citep{Ridge06,Pineda08}, with outflows identified
across the entire cloud in an unbiassed manner using the
\twelcons(1--0) dataset \citep{Arce10},
and outflows around the
dense cores mapped in $^{12}$CO(3--2) \citep{Hatchell07b}.

Since a wealth of data exist for the Perseus molecular
cloud, it is possible to move beyond characterizing properties
of structures in the cloud based on a single tracer, and advance
to a study of the kinematic relationship between the structures on 
different scales within the cloud.

Several studies have already been carried out which compare the column
density features on various scales in Perseus.  \citet{Jorgensen07}
related the young embedded protostars to the SCUBA dense cores
that they inhabit and showed that the protostars have 
little, if any, motion within the dense core they form out of.
In turn, the SCUBA dense cores are found only within
high extinction zones within the molecular cloud, suggesting
an extinction threshold of A$_V \sim 5$~mag 
for dense core formation \citepalias{Kirk06}.
Such a threshold is consistent with predictions from a 
magnetically-dominated formation scenario, wherein dense
core formation is only possible in the highest column
density regions, where the ionization fraction is low enough
to allow for a fast ambipolar diffusion timescale \citep{McKee89}.
The dense cores also tend to be offset from the peaks of
the larger extinction regions that they inhabit in a pattern
suggestive of a triggered formation scenario \citepalias{Kirk06}.  A nearby
B0 star, 40 Persei, previously suggested as a trigger for 
star formation in the nearby L1451 region \citep{Walawender04},
is located in a position to be consistent with this
scenario \citepalias{Kirk06}.

A multi-scale analysis of kinematics comparable to the column
density studies has not yet been
performed, although the previous single-scale analyses reveal
information which already places constraints on the
star formation scenario.
The pointed \nhns(1--0) and \cons(2--1) observations \citepalias{Kirk07} 
of the dense cores showed
that the dense cores in Perseus are intimately tied to their
surrounding envelope of material.  The velocity
dispersion of each core's 
\nh gas tends to be thermally-dominated, while
the velocity dispersion of each core's surrounding \co 
gas tends to be
much more non-thermal.  Despite this, the difference in centroid
velocities between the \nh and \co gas is typically less 
than the sound speed, and much less 
than the velocity dispersion of the \co gas. 
Similar results were found from the pointed NH$_3$(1,1) and (2,2) and CCS(2--1)
survey \citep{Rosolowsky08}.  {\hknew In Perseus, \nh and NH$_3$ have
been shown to trace each other very well \citep{Johnstone10}.}

In a survey of isolated dense
cores in a variety of star-forming regions,
\citet{Walsh04} found similarly small velocity differences
and argued that this
is evidence against the competitive accretion scenario.
\citet{Ayliffe07} analyzed a simulation where cores undergo
competitive accretion and produced mock observations to match
the \citet{Walsh04} survey.  \citet{Ayliffe07} found they could match the
small velocity differences at later times in the simulation, 
however, the dense core
line width at later times greatly exceeded that found in
observations.
This illustrates that `observations'
of simulations are necessary for comparisons, as 
they can produce unexpected results. 
The Perseus 
observations of \citetalias{Kirk07} extend the result
of small centroid velocity differences
to a more clustered star formation environment, where
competitive accretion is predicted to be more effective
\citep[see, for example,][]{Ayliffe07}.

Using estimates of the mass
and size of the dense cores from the SCUBA observations
and the internal velocity dispersion from the \nh survey,
\citetalias{Kirk07} 
furthermore showed that each dense core appears to be
in approximate ``virial equilibrium'', so long as the surface pressure
on the cores (from the overlying mass of the cloud) is
considered.

Simulators have started to `observe' similar kinematic measures
in their simulations, such as the internal velocity
dispersion of cores (Klessen et al.~2005; Ayliffe et al.~2007;
Offner et al.~2008; Kirk, Johnstone, \& Basu 2009, hereafter KJB09)
and the core-to-envelope velocity difference 
(Ayliffe et al.~2007; KJB09).
These measurements allow for a direct comparison between the 
observational kinematic surveys and the simulations.
In this paper, we use the multi-scale kinematic data available in
the Perseus molecular cloud to provide further observational
measures by which
to test future models of star formation.
{\hknew In the interest of facilitating the comparison between
observations and simulations, we focus on physically-motivated
quantities that are simple to derive.  It is hoped that these simple
observational measures can be related to simulations without many
additional assumptions required for the simulations.}

In Section~2, we summarize the observational data used
in our analysis, while in Sections 3-6, we present results based
on small- (Section~3) to large- (Sections~5 and 6) scale quantities.
Interpretation of the results is discussed in Section~7, and we
conclude in Section~8.  A glossary of the kinematic measures used
in this paper is provided in Appendix~A.


\section{OBSERVATIONS}
Our analysis here is based primarily on two datasets - a pointed survey of
the kinematics of the dense cores (\nhns) and their surrounding
material (\cons) in Perseus \citepalias{Kirk07}, and a survey of the 
less-dense gas (\thircons) spanning the entire Perseus cloud
\citep{Ridge06,Pineda08}.  Additionally, we make use of the large
structures (termed `extinction regions' in this paper) identified in an 
extinction map of Perseus based on analysis of the 2MASS dataset
\citepalias[`extinction super cores' in][]{Kirk06}.  
Here, we summarize the relevant information
for each of these datasets.

\subsection{Pointed \nh and \co Observations}
Using the IRAM~30~m telescope, \citetalias{Kirk07} made single pointing
observations in frequency-switching mode
of \nhns(1--0) and \cons(2--1) in 157 candidate 
dense cores in Perseus.
The beamsize was 25\arcsec\ for \nh and 11\arcsec\ for \cons.
The positions observed were selected primarily based on the SCUBA 
submillimetre analysis of \citetalias{Kirk06}, with additional 
targets selected based on a visual inspection of the Palomar plates
and the location of the maxima of large-scale structures seen
in the 2MASS-based extinction map of Perseus \citepalias{Kirk06}.
\nh emission was detected in the majority of pointings (97), 
with the highest
fraction of detections belonging to the SCUBA dense core list 
(84\%) and lower fractions (42\% and 14\% respectively)
for the others.  \co emission was detected in all but one
of the pointings.  
Linewidths, centroid
velocities, and integrated intensities were measured 
for all of the spectra with detected signal.
\nhns(1--0) has seven hyperfine components in its spectrum
and these were simultaneously fit using CLASS's hyperfine
fitting program.  The vast majority of the \nh spectra
required a single velocity fit,
and of the few cases where a
second velocity component was required, the two components
were usually discernible in the spectrum.  The \co spectra
were fit with a single Gaussian where appropriate, but
more frequently required a two-component Gaussian fit.  
Table~\ref{tab_rel_motions} summarizes the relevant results
used in this analysis: the difference in centroid velocities,
$v_{N2H+} - v_{C18O}$, 
and the linewidth (one sigma of a Gaussian fit) 
of the \cons, $\sigma_{C18O}$, for every \nh 
detection\footnote{A summary of all of the kinematic measures
and their associated symbols used in this paper can be
found in Table~5.}. 
See Tables~3 and 4 in \citetalias{Kirk07} 
for the full spectral fit parameters for \nh and \co respectively.
Note that when \nh spectra are plotted in this paper, the
single Gaussian profile derived from the model fit is shown for simplicity.

The dense cores were further
divided into protostellar and starless, based on associations with
the Perseus embedded YSO catalog of
\citet{Jorgensen07}.  This YSO catalog uses two criteria of which protostars
must satisfy at least one.  The first criterion
is the spatial coincidence of a {\it Spitzer} source and a SCUBA core,
and the second criterion 
is colours that satisfy [3.6] - [4.5] $>$ 1 and
[8.0] - [24] $>$ 4.5 in the {\it Spitzer} bands, i.e., colours
that have been previously shown to select YSOs
\citep[see][for more details]{Jorgensen07}. 

\subsection{\thirco Map}
The \thircons(1--0) data were obtained by the COMPLETE 
team\footnote{All of the COordinated Molecular Probe Line Extinction Thermal  
Emission (COMPLETE) Survey data are publicly available at
\url{http://www.cfa.harvard.edu/COMPLETE/ }}
using the SEQUOIA focal plane array at the FCRAO telescope \citep{Ridge06}.  
The region mapped covers $\sim$20~$\deg^2$, with a  
46\arcsec\ angular resolution and 0.066~km~s$^{-1}$ velocity channels.  
The data were baseline-subtracted and {\hknew sampled} onto a 23\arcsec\ grid;
the average noise is 0.1~K (in $T_A^*$ scale).
Detailed comparisons between  
the molecular line data and extinction are presented in \citet{Pineda08}
and \citet{Goodman09}.


In our comparison of the \nhns, \cons, and \thirco spectra
at the position of each core (Section 3), we fit a single Gaussian 
to each relevant \thirco spectrum.  The difference in \nh and \thirco
velocity centroids and \thirco velocity dispersions
are given in Table~\ref{tab_rel_motions}.  
All fits were visually examined; poor
fits (i.e., where the spectra are not well-described by a single
Gaussian) are also noted in Table~1.  The estimation of the
kinematic properties of the larger extinction regions using the
\thirco data are discussed in Section~4.

\subsection{2MASS Extinction}
{\hknew Extinction maps are now readily available for the entire
sky \citep[see][]{Rowles09}, allowing large-scale structures to be
easily identified in all molecular clouds.}
A map of the extinction
throughout the Perseus cloud has been produced 
based on reddening of stars in the
2MASS catalog (Ridge et al.~2006; KJD06). 
This map was created using
the NICER technique \citep{Lombardi01}, and was presented in 
\citetalias{Kirk06} and \citet{Ridge06}.  
The extinction map provides a useful dataset for
defining the large-scale structures in Perseus.  While this is not
the only way to classify the large-scale column density 
structures, it has the advantage of
being simpler than disentangling structures in a spectral cube.
In datasets with clustered emission features, which are
particularly prevalent in \thirco maps of molecular clouds
(and generally more common in 3D datasets),
the emergent properties of the structures 
identified are particularly sensitive to the precise structure-identifying 
algorithm and parameters used \citep[see, for example,][]{Pineda09}.
For our analysis here, we use the large-scale structures identified
in \citetalias{Kirk06} using the 2D version of the 
{\it clfind} \citep{Williams94}
algorithm with a threshold of 3~mag and a stepsize of 1~mag 
on the extinction map smoothed to 5\arcmin\ resolution.
The relevant properties of these extinction regions are summarized in
\citetalias{Kirk06} (Table~3) and \citetalias{Kirk07} (Table~5).
{\hknew As a test, we also ran our analysis using the independently-defined 
large-scale regions in \citet{Pineda08} based on CO centroid velocities 
and dust temperatures derived from IRAS data.  Our conclusions are
unchanged using these alternate regions, suggesting that our results
are not strongly dependent on the definition of the regions used.
We prefer to define the regions based on extinction, however, as this
definition should be easy to apply to other observations and simulations.}

\begin{figure}[tp]
\includegraphics[angle=90,height=18cm]{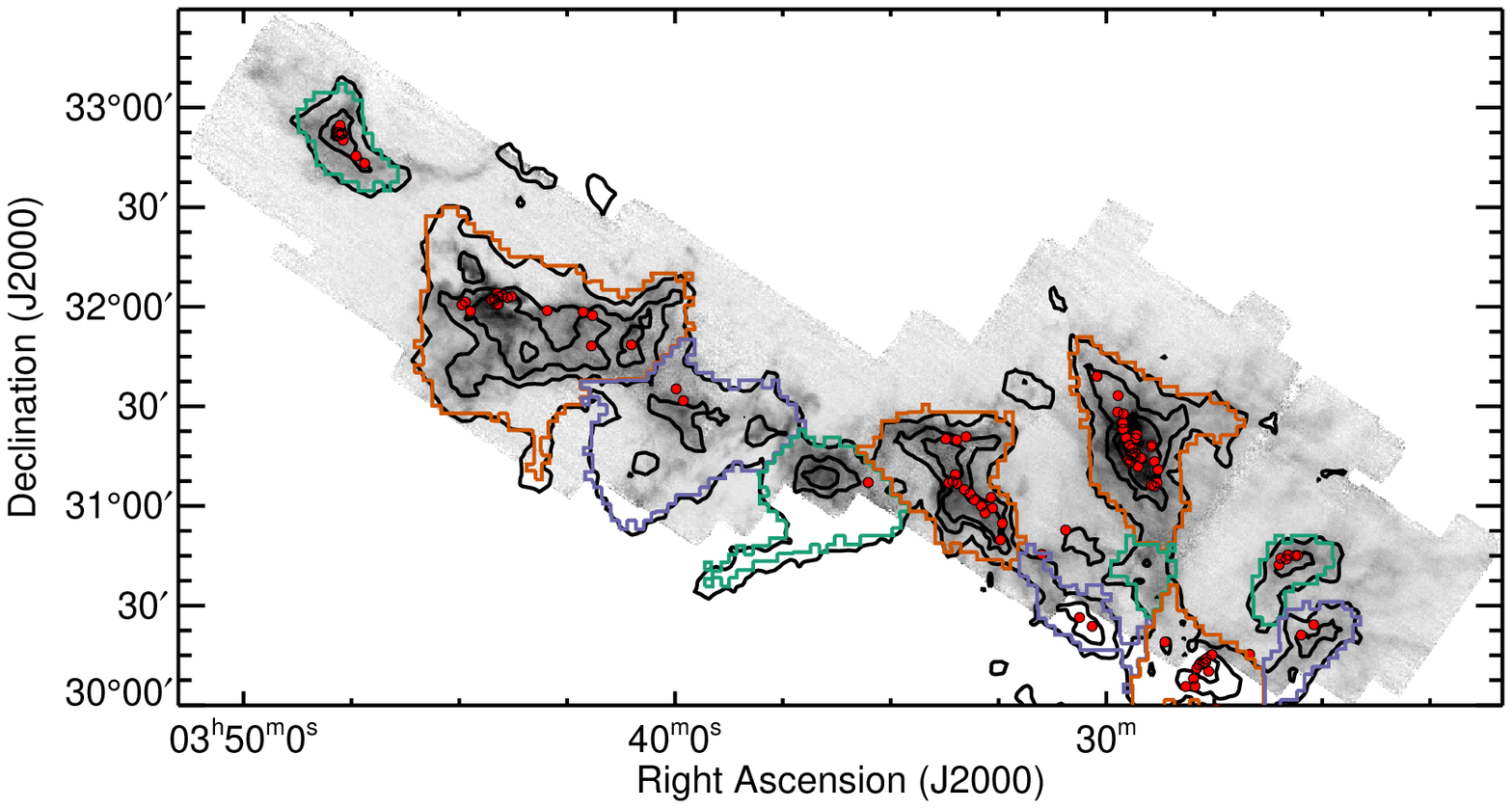}
\caption{The greyscale image shows the \thirco integrated intensity map from
	the FCRAO telescope.  The overlaid black 
	contours show extinction
	measurements derived from 2MASS data at levels of
	A$_V$ = 3, 5, and 7 mag.  The
	red circles show the locations of all of the IRAM~30~m
	\nh observations.  The extinction regions used in the analysis
	here are denoted by the coloured contours. See text for details.
	}
\label{thirco_plus_targets}
\end{figure}

Figure~\ref{thirco_plus_targets} summarizes the data analyzed in this
paper.  The greyscale image shows the integrated intensity of
the \thirco data, while the red circles show the \nh pointings.
The total column density, as measured by the extinction observed
in the 2MASS dataset is overlaid as black contours and the extinction
regions are shown as coloured contours.

\section{RELATIVE MOTIONS AT THE DENSE CORE POSITIONS}
\citetalias{Kirk07} measured the difference in centroid
velocity of \nh and \co ($v_{N2H+} - v_{C18O}$), reflecting
the motion between the dense cores (\nhns) and their surrounding
envelopes (\cons) along the line of sight.  This relative motion 
was found to be smaller than the envelope velocity
dispersion ($\sigma_{C18O}$) and for 90\% of the cores, 
less than the sound speed, assuming a temperature of 
15~K\footnote{In terms of the full three-dimensional picture, 
the full (3D) velocity difference between the core and envelope 
would be expected to be larger than we observe, 
since only the LOS components of the velocities are measurable.  
Similarly, the envelope velocity dispersion
and sound speed discussed also only reflect LOS components of
the full 3D motion; all are equally affected by projection.}.
The small difference in centroid velocities 
implies that the dense core is dynamically
coupled to its surrounding envelope, and appears to argue
against the competitive accretion scenario 
(KJT07; Walsh et al~2004, but see also Ayliffe et al~2007).
Since \thircons(1--0) is sensitive to lower density material than
\cons(2--1), we can additionally compare the \nh and \co centroid
velocities to that of the \thirco to determine whether the 
tight correlation in velocities extends to the less dense
material.  (Note that the beamsize of the \thirco observations
is several times larger than that of the other tracers, so
a larger physical volume of material is being traced, 
both in extent across the plane of the sky and along the LOS.)

\begin{figure}[p]
\includegraphics[height=14cm]{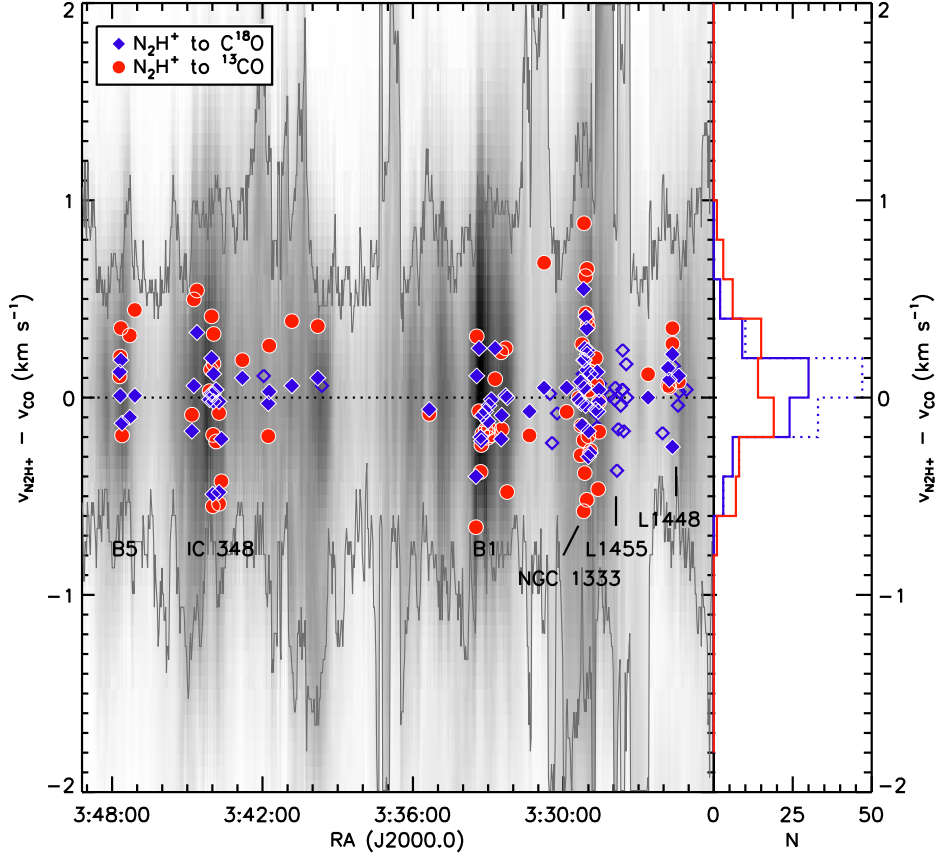}
\caption{
	The distribution of centroid velocity difference between \nhns,
	\cons, and \thirco as a function of RA (left panel), and the
	full distribution (right panel).  In the left panel,
	the blue diamonds indicate
	the distribution of $v_{N2H+} - v_{C18O}$, while the red
	circles indicate the distribution of $v_{N2H+} - v_{13CO}$.
	The empty diamonds indicate locations where there is no 
	good Gaussian fit to the \thirco data (or \thirco observations 
	do not exist), while the filled diamonds indicate where
	there is a good Gaussian fit to the \thirco data.
	The greyscale image shows the position-velocity 
	diagram for the \thirco gas {\hknew and the grey contours show
	the half-maximum value at every RA}.  
	At each RA, the cumulative spectrum
	has been shifed so that the peak intensity falls at a value
	of zero on the diagram.  The greyscale ranges from 
	2~K~km~s$^{-1}$ (black) to 0~K~km~s$^{-1}$ (white).
	The approximate RA of well-known star forming regions are
	indicated by the text.  In the right panel, the dotted blue line
	indicates the full distribution of $v_{N2H+} - v_{C18O}$, while
	the solid blue line shows only the points where there is a good
	Gaussian fit to the \thirco data.  The distribution of
	$v_{N2H+} - v_{13CO}$ is shown in red.
	}
\label{fig_nh_vs_co_point}
\end{figure}

Figure~\ref{fig_nh_vs_co_point} shows the distribution
of centroid velocity differences as a function of RA (left panel)
and the overall distribution of values (right panel).
The filled symbols (left) and solid lines (right) denote locations where
the \thirco Gaussian fits were deemed `good' (roughly 87\%),
while the empty symbols and dotted lines denote locations where
only the \nh and \co fits were good.  Note that the \thirco map
did not extend to L1455, hence no `good' fits in that region were possible.
All of these data are also provided in Table~\ref{tab_rel_motions}. 
The greyscale image in the figure background shows the position-velocity
(PV) diagram for the \thirco gas, {\hknew with the contours indicating
the FWHM values}.  
As can be seen from the figure, the spread
in velocity difference is much smaller between \nh and \co
than between \nh and \thircons, and both of these are smaller
than the range spanned by \thirco emission at that RA.
The standard deviation of the centroid velocity differences
are 0.18~km~s$^{-1}$ for $v_{N2H+}-v_{C18O}$ at all locations, 
0.19~km~s$^{-1}$ for  $v_{N2H+}-v_{C18O}$ where good \thirco fits exist,
and 0.33~km~s$^{-1}$ for $v_{N2H+}-v_{13CO}$ {\hknew , while the
errors in the centroid velocity differences measured are
roughly 0.015~km~s$^{-1}$ for \nh and either CO species.
A two-sided KS test\footnote{A two-sided KS test allows one to measure 
the probability of two distributions being drawn from the same parent 
distribution, \citep[e.g.,][]{Lupton93}.} 
shows that the likelihood that both distributions 
are drawn from the same parent sample is only 2\%.} 
The sound speed
is 0.23~km~s$^{-1}$ for the mean gas, assuming a temperature
of 15~K and a mean molecular weight of $\mu = 2.33$.  
The largest velocity differences are found in the
NGC~1333 region.

\begin{figure}[hp]
\plotone{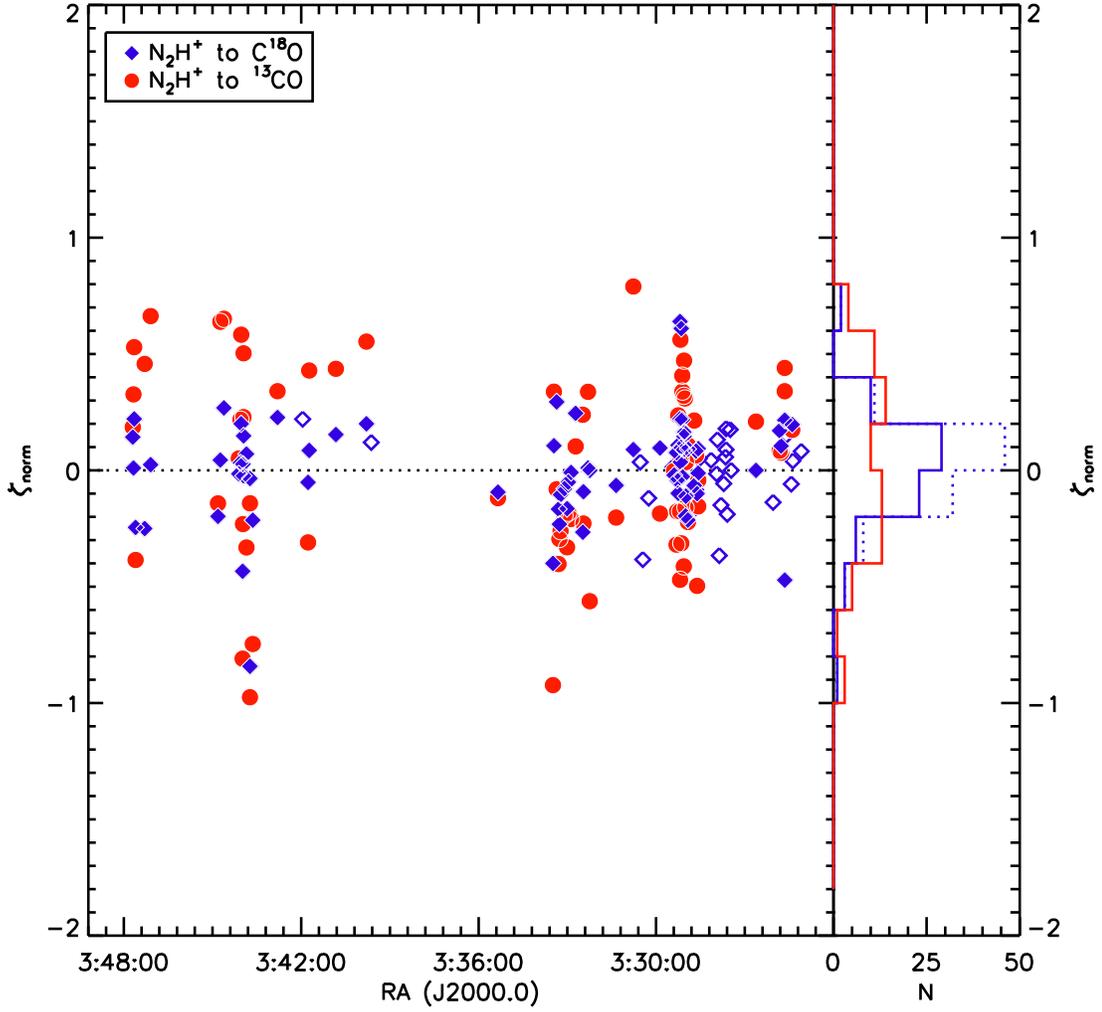}
\caption{
The distribution of normalized velocity differences, $\zeta_{norm}$,
	between \nhns, \cons, and \thirco as a function of RA (left panel).
	The total distribution is shown in the right panel.  See
	Figure~\ref{fig_nh_vs_co_point} for details on the plotting 
	conventions used.
}
\label{fig_nh_vs_co_point_norm}
\end{figure}

A second way to measure the relative motion between the species
is to compare the difference in velocity to the velocity dispersion
of the less dense species.  This normalized velocity difference
is written as:
\begin{equation}
 \varsigma_{norm} = \frac{v_{N2H+} - v_{CO}}{\sigma_{CO}}
\end{equation}  
This quantity indicates whether
the dense core moves within the typical range of motions present in the 
lower density environment, in this case, characterized by the
gas along the same line of sight as the core.  
\citetalias{Kirk09} showed that the
normalized velocity difference can provide strong discrimination between
numerical simulations with varying input levels of turbulence and
magnetic field strengths.  The normalized velocity difference was also
one of the most discrepant measurements between many of the simulations
analyzed in \citetalias{Kirk09} and the observations of \citetalias{Kirk07}.
Figure~\ref{fig_nh_vs_co_point_norm} shows the normalized velocity
difference between the three species, as a function of RA
(left panel) as well as the full distribution (right panel).  As with the
centroid velocity difference, we find the normalized velocity
difference also tends to be quite small -- nearly always less than
one. 
The standard deviations of the normalized velocity differences
are 0.20 (\nh - \co at all positions), 0.22 (\nh - \co
where good \thirco fits exist), 
and 0.39 (\nh - \thircons){\hknew, while the errors are 0.014 and
0.019 for \nh to \co (all positions) and \nh to \thirco
respectively.  A two-sided KS test shows that the likelihood that 
both distributions are drawn from the same parent sample is 0.2\%.}
This indicates that the motions of the dense and less-dense
gas are correlated (i.e., closer than expected from a random
distribution), 
but less so at larger scales.
This is perhaps not surprising, since the \thirco
observations trace not only a larger scale across the plane
of the sky (due to the larger beamsize), but also are expected
to trace a much larger scale along the LOS, and likely include
emission from additional lower-density structures not associated
with the dense core, due to \thircons's sensitivity
to lower density material than \cons.
{\hknew Comparison with the results in Section~4.2 suggests that
the agreement between \nh and \thirco centroid velocities might
become slightly stronger if the \thirco beamsize was small enough
to match the \nh beam.  We expect, however, that the bulk of the
difference is due to the larger scale of material traced by the \thirco along
the LOS, which would be unchanged with a smaller beamsize.}

A slight skew is apparent in both \nh - \thirco distributions,
{\hknew but does not appear to be physical.
The mean velocity difference between \nh and \thirco is
0.03~km~s$^{-1}$, less than half the size of the \thirco spectral
resolution, and hence is likely due to stochastic effects.  As
a second check, we examined the velocity differences within each
extinction region and found that there was no systematic offset;
offsets of opposite directions were found in different regions.}


\section{DENSE CORES RELATIVE TO THE EXTINCTION REGIONS}
\subsection{Calculating Extinction Region Properties}
\begin{figure}[hp]
\plotone{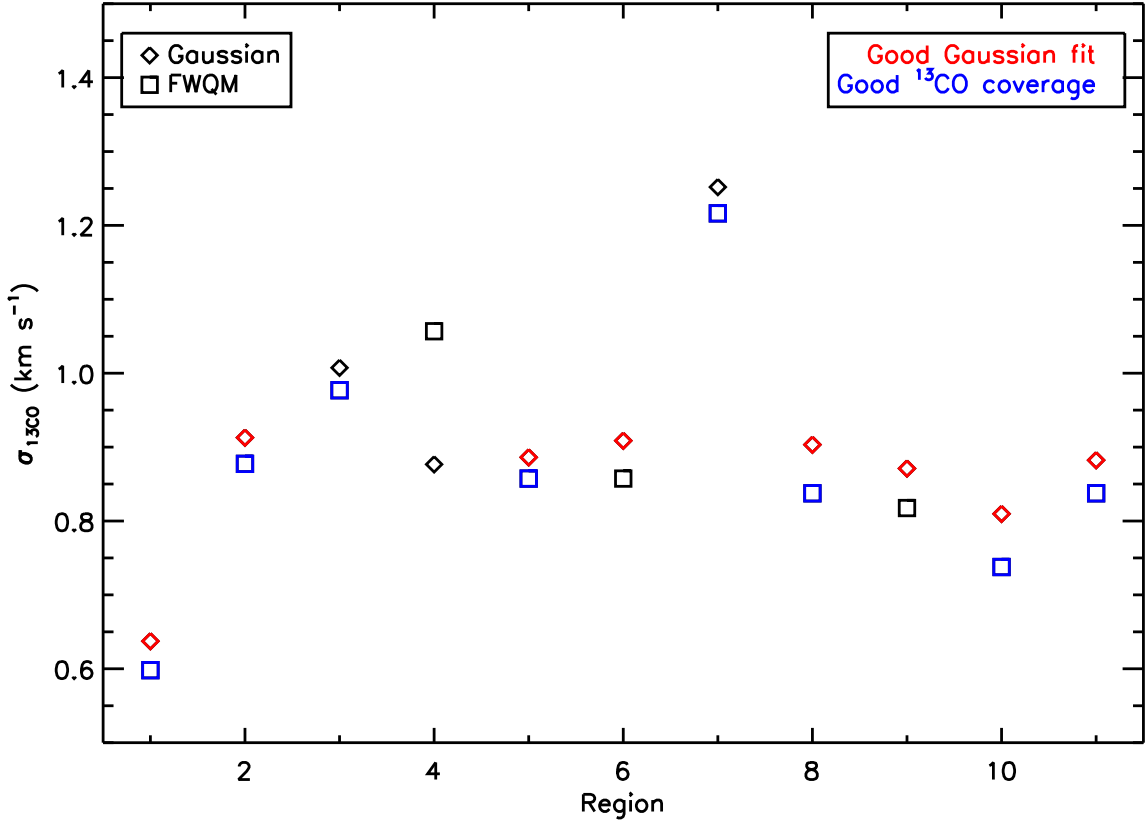}
\caption{A comparison of the different measures of the velocity
	dispersion within each extinction region, illustrating that
	the measures are similar in regions where the \thirco coverage is 
	good. 
	The \thirco velocity dispersion is measured with both a single
	Gaussian (diamonds) and the equivalent
        Gaussian sigma derived from measuring the full width at
        quarter maximum (FWQM) of the spectra (squares).
	Red points denote the single Gaussian fits that were judged
	by eye to fit well, while blue points denote regions where 
	the \thirco data covers 80\% or more of the extinction region, 
	i.e., the spectrum is a good representation of that for the 
	entire region. In all regions where there is good \thirco coverage
	(blue points), both velocity dispersion measurements
	(diamonds and squares) are similar, regardless
	of whether the single Gaussian fits were judged to be
	good (red points) or not.  The remainder of this paper uses the
	single Gaussian fit measures, restricted to regions where there is 
	good \thirco coverage.
	}
\label{fig_disp_measures}
\end{figure}

Since a full map exists for the \thirco data, it is also
possible to quantify the motions of the dense cores within
the larger extinction regions identified in \citetalias{Kirk06}.
In order to more easily identify large-scale features
in the regions, 
we first convolve and re-grid the \thirco map
to match the 2MASS extinction map angular resolution
(5\arcmin) and grid.  We 
create the mean \thirco spectrum of the extinction region
by summing all \thirco spectra that lie within the contours of the
extinction region boundaries of \citetalias{Kirk06}.  We then calculate
the centroid velocity, $v_{13CO, reg}$, and velocity dispersion,
$\sigma_{reg, Gauss}$, of the spectrum using a single
Gaussian fit.  
{\hknew Note that these spectra are little affected by the prior
map convolution to 5\arcmin -- in most cases, there is less than 3\% variation
between the Gaussian fit parameters for the total spectra of the
convolved and un-convolved maps.}
Several of the spectra are not well-fit by
a single Gaussian; in \citetalias{Kirk09}, we found that the full
width of emission was better characterized by 
the full width at quarter maximum (FWQM).
To determine the reliability of the Gaussian
fits to the spectra, we also measure the FWQM for the \thirco
region spectra, following the procedure outlined in
\citetalias{Kirk09}, and 
{\hknew calculate $\sigma_{reg, FWQM}$, the sigma (1/e width) that 
would be measured if the spectrum were a perfect Gaussian with
the measured FWQM.}
Table~\ref{tab_reg_widths}
provides the values fit to the cumulative spectra for each
region using both methods, while Figure~\ref{fig_disp_measures}
shows both results for all of the regions.
Three of the eleven extinction regions 
have significantly incomplete \thirco coverage.  Regions that
are not well-covered may not provide a good representation of the
overall region kinematics, regardless of the method used to
fit the total spectrum.  We set a minimum value of 80\%
coverage in the \thirco observations for all of our subsequent
analysis.  Figure~\ref{fig_disp_measures} shows that
when a cutoff of 80\% is applied, both measures of the total velocity
dispersion ($\sigma_{reg,Gauss}$ or $\sigma_{reg,FWQM}$) 
produce similar results.  For the remainder of the
analysis and for the normalized velocity difference between
the \nh cores and their associated extinction regions 
provided in Table~\ref{tab_rel_motions},
only the Gaussian fit results are used.
{\hknew Note that the tendency for the $\sigma_{reg,FWQM}$ measures
to be slightly lower than the corresponding $\sigma_{reg,Gauss}$
value appears to be primarily caused by our limited resolution.
While the FWQM is a good measure for identifying fainter, broad components
of spectral lines, when significant line wings are not present,
noise and the small number of spectral channels will tend to
make the FWQM underestimate the true width, as this measure
is not interpolated across spectral channels (unlike the Gaussian fit).}

\subsection{Analysis}
\begin{figure}[hp]
\plotone{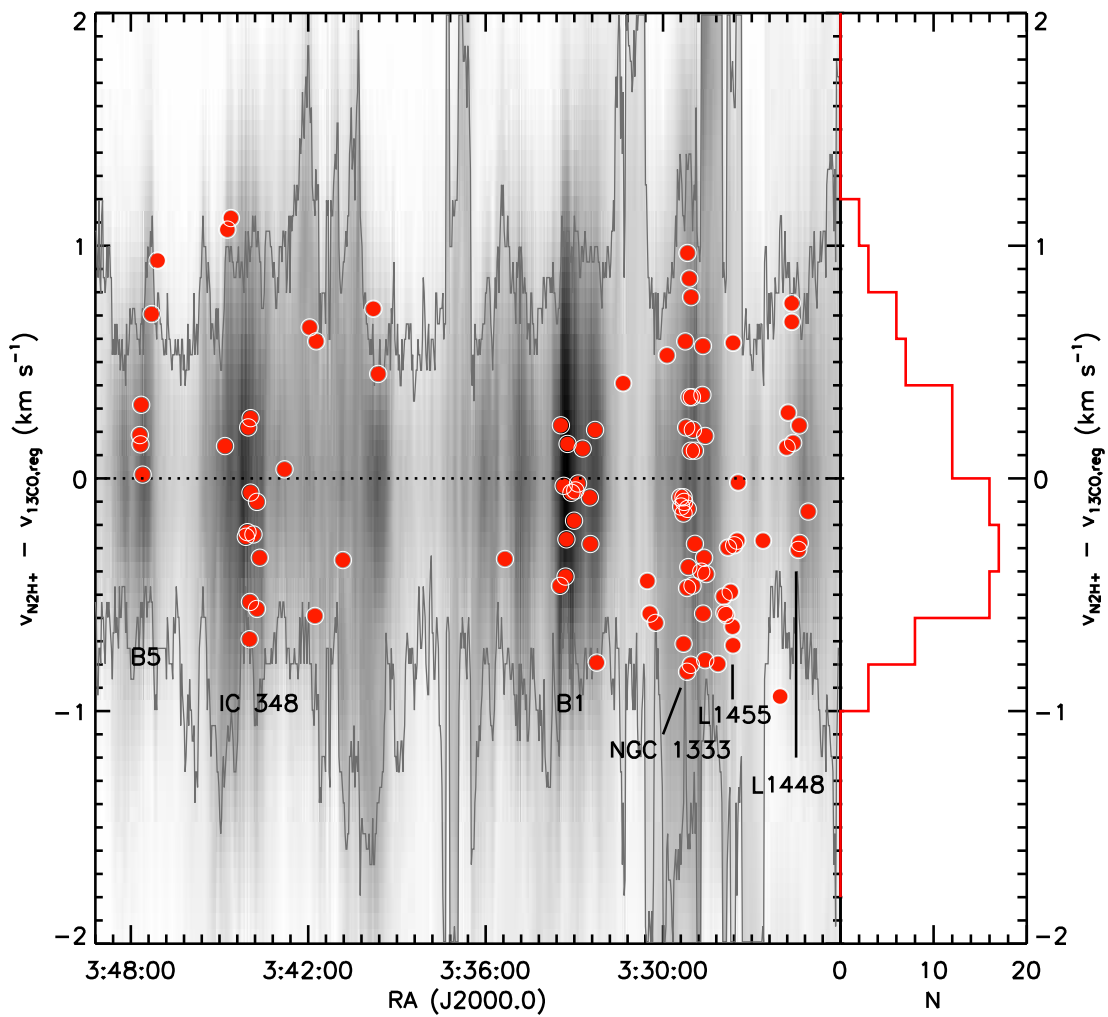}
\caption{
The distribution of the centroid velocity difference between
\nh and the \thirco region in which the core is located as a function
of RA (left panel), and the total distribution (right panel).
	The greyscale image in the left panel shows the position-velocity (PV)
	diagram for the \thirco gas {\hknew with the grey contours
	showing the half-maximum value at every RA}.  
	At each RA, the cumulative spectrum
	has been shifed so that the peak intensity falls at a value
	of zero on the diagram.  The greyscale ranges from 
	2~K~km~s$^{-1}$ (black) to 0~K~km~s$^{-1}$ (white).
}
\label{fig_core_vs_region}
\end{figure}

\begin{figure}[hp]
\plotone{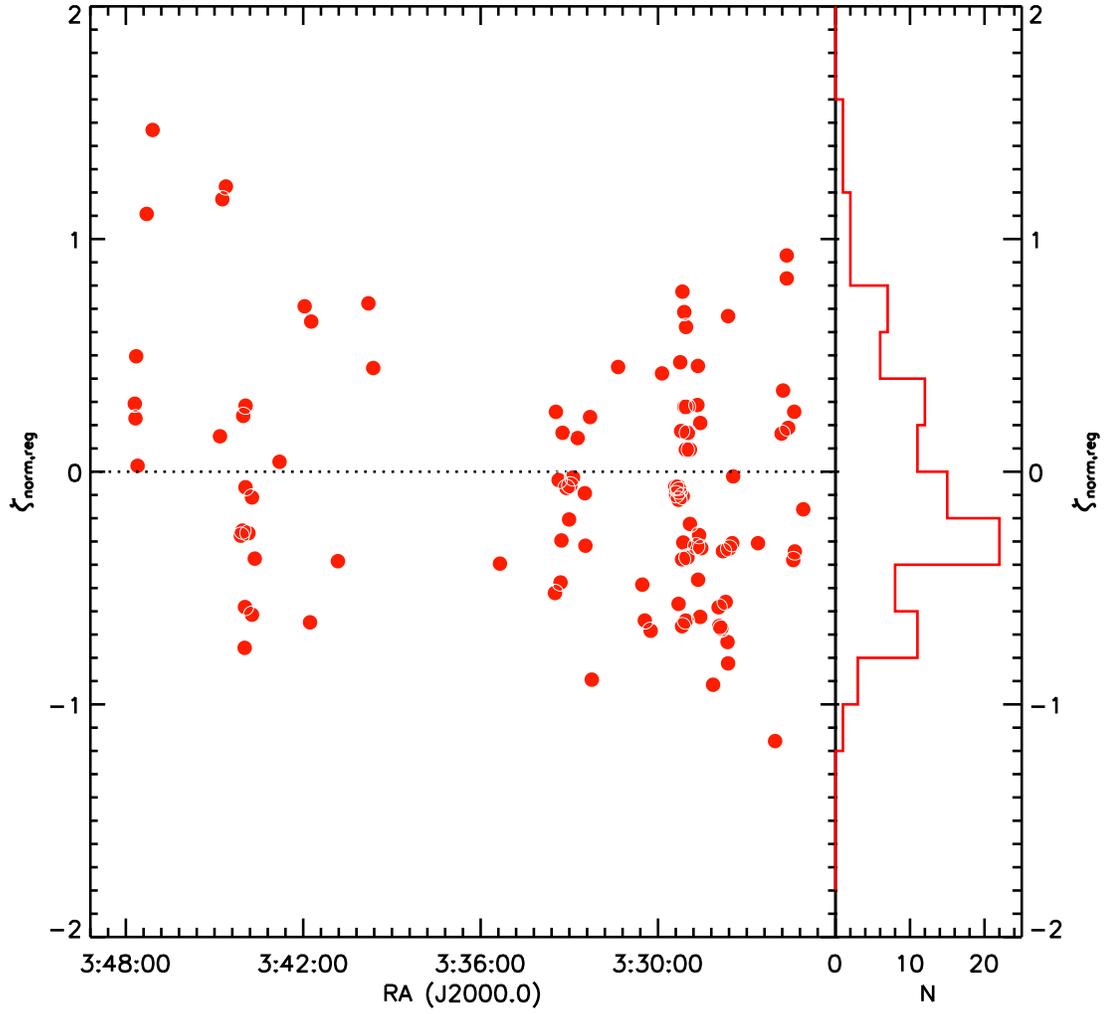}
\caption {
The distribution of normalized centroid velocity differences,
$\zeta_{norm}$ between the \nh and the region in which the core is
located as a function of RA (left panel) and the total distribution
(right panel).  See previous figure for the plotting conventions used.
}
\label{fig_core_vs_region2}
\end{figure}

Figure~\ref{fig_core_vs_region} shows the distribution
of the centroid velocity differences between the \nh and the
\thirco region, $v_{N2H+} - v_{13CO,reg}$ versus RA (left
panel), in addition to the full distribution of velocity
differences (right panel).  
The motions between the core and region again tend to be small,
although larger than the difference using the \thirco velocity
at the location of the dense core.  The standard deviation
of the distribution is 0.48~km~s$^{-1}$ (c.f. 0.33~km~s$^{-1}$
for \thirco gas at the dense core location){\hknew; the error
in the velocity difference measured is roughly 0.011~km~s$^{-1}$.
Using a two-sided KS test shows that the probability of the
distribution of velocity differences between the core and the \thirco
gas at the core's location versus over the entire region is only 0.5\%.}
Figure~\ref{fig_core_vs_region2} shows the distribution of 
normalized velocity difference between the cores and regions,
$\zeta_{norm,reg}$ (calculated using eqn 1),
 versus RA (left panel) as well at the total
distribution (right panel).  The standard deviation of $\zeta_{norm,reg}$
is 0.5 (c.f. 0.39 for \thirco gas at the dense core location)
{\hknew; the error in the normalized velocity difference measured
is 0.013.  The two-sided KS test gives a probability that this normalized
velocity difference and the one measured for the \thirco at the core's
location are drawn from the same parent distribution of 7\%.}
{\hknew We also examined the relationship between the velocity difference
and the separation from the centre of the region, but found
no discernable trend.}

\section{CORE KINEMATICS PER REGION}
Most of the extinction regions contain multiple dense cores, thus
the relationship between the ensemble of dense cores and the
extinction region can also be examined.  
In particular, the core-to-core (v$_{LSR}$) 
motions can be compared to
the spread of motions seen over the region as a whole
in an ``environmental'' tracer (e.g., \thircons);
this should reveal
how connected the dense cores are to the lower density gas
in the region.
If the dense cores are connected
to the large-scale non-thermal motions, then the dispersion in
the core centroid velocities would be expected to be similar to the
global velocity dispersion of the region, as measured in \thircons.  
If, however,
the core-to-core motions are much smaller than the global
velocity dispersion of the region, then this may be an
indication that the dense cores have become detached
from the large-scale motions within the cloud, arising from
turbulent flows, for example.
Alternatively, the spatial distribution of cores may lead
to them tracing a smaller volume of material than the \thircons.

\subsection{Method of Calculation}
Section~4.1 describes the calculation of the extinction 
region velocity dispersions.
There are several approaches that can be taken to quantify the
core-to-core motions within each extinction region.  The first,
method A, is a purely statistical measure, which involves
simply taking the standard
deviation of the centroid velocities of all of the dense cores
within each region, $\sigma v_{meth. A}$.  
This is effective when the spectrum of
each core is well described by a single, relatively narrow,
velocity component.  In cases where these conditions do not hold,
interpretation of the standard deviation can be more difficult.
This method also does not account for individual core linewidths,
which could represent a significant fraction of the total
dense gas velocity dispersion of the core gas. 
The second method, method B, accounts for both the core-to-core
motion and the internal velocity dispersion through summing
all of the spectra and measuring the resultant total
velocity dispersion, $\sigma v_{meth. B}$.  
This method may be biassed if some cores
are much brighter than others, and can be particularly
challenging for molecules with hyperfine transitions, due to
different components blending together in the summed spectrum.
In \citetalias{Kirk07}, we adopted
method A for measuring the \nh relative motions and 
method B for measuring the \co relative motions; here, we again adopt
method A for \nhns, but show that method B yields similar
results.  Note that any velocity gradient across the region 
is implicitly included in both measures and is
not accounted for separately; the same is true for the extinction
region veloctiy dispersion.

\subsection{Results}
\begin{figure}[hp]
\plotone{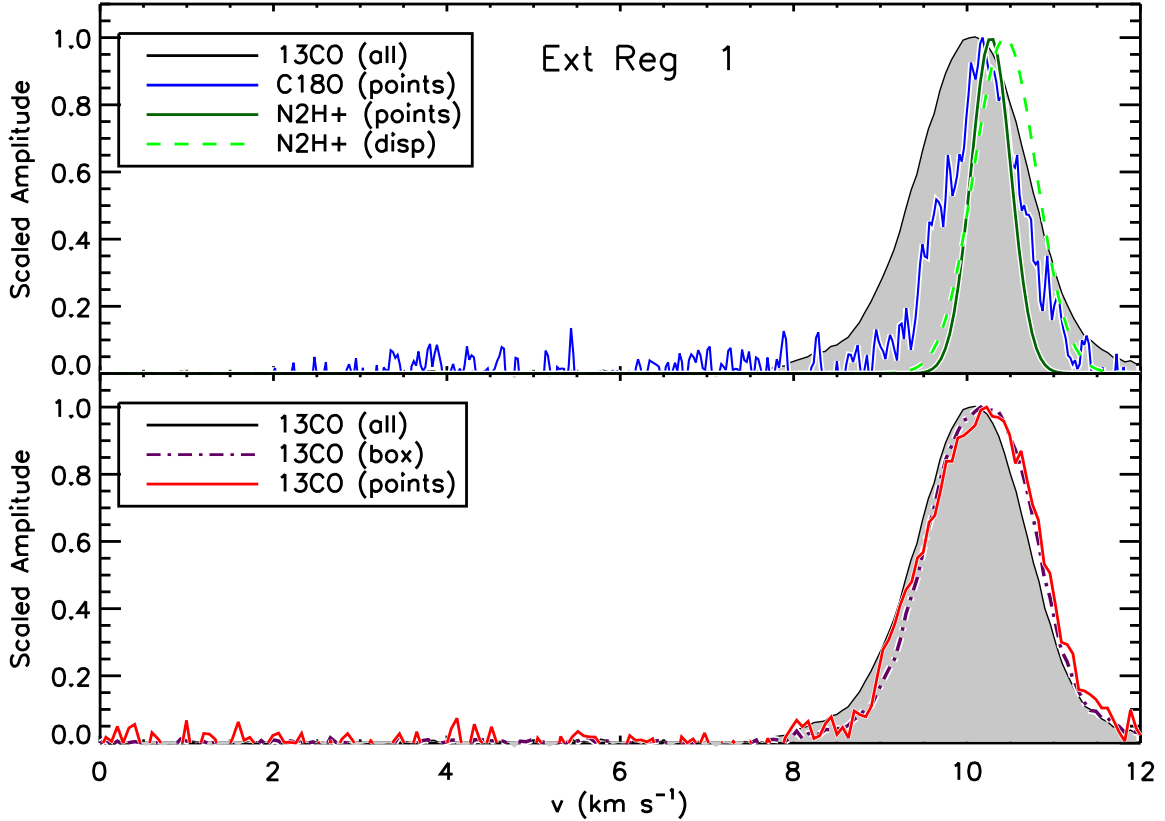}
\caption{A comparison of cumulative spectra within extinction region 1.  
	Top panel: The \thirco spectrum of the entire region from 
	the top panel (black line and grey shading), 
	the \co spectrum summed over all locations in the region with
	\co detections (``method B'', blue line), the resultant 
	fit for a similar summation 
	for \nh (``method B'', dark green line), and the standard deviation of
	the centroid \nh velocities in each region (``method A''), 
	centered on the
	mean value (light green dashed line).
	Bottom panel: \thirco spectra
	derived from summing the spectra over the entire region (black line),
	summing over the box in which the \nh dense cores are found only
	(purple dot-dashed line), and summing over only the points where
	where \nh dense cores were detected (red line).  
	Note the \thirco spectral data has a full range of
	$\sim -10$ to 30~km~s$^{-1}$; the truncated range of 0 to 
	12~km~s$^{-1}$ plotted covers the range of velocities
	where \nh and \co emission was observed.
	}
\label{fig_compare_spec_1}
\end{figure}

\begin{figure}[hp]
\plottwo{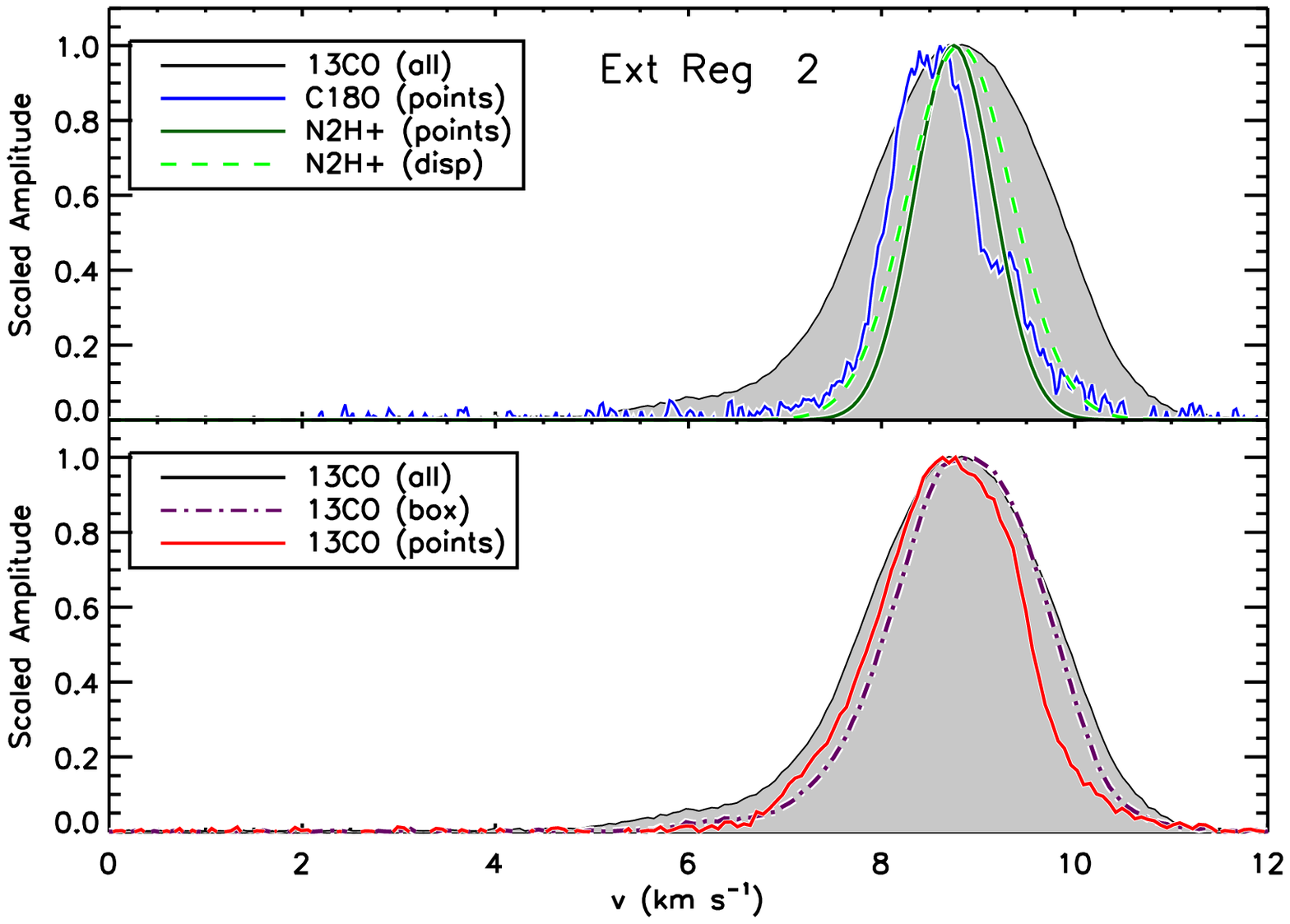}
	{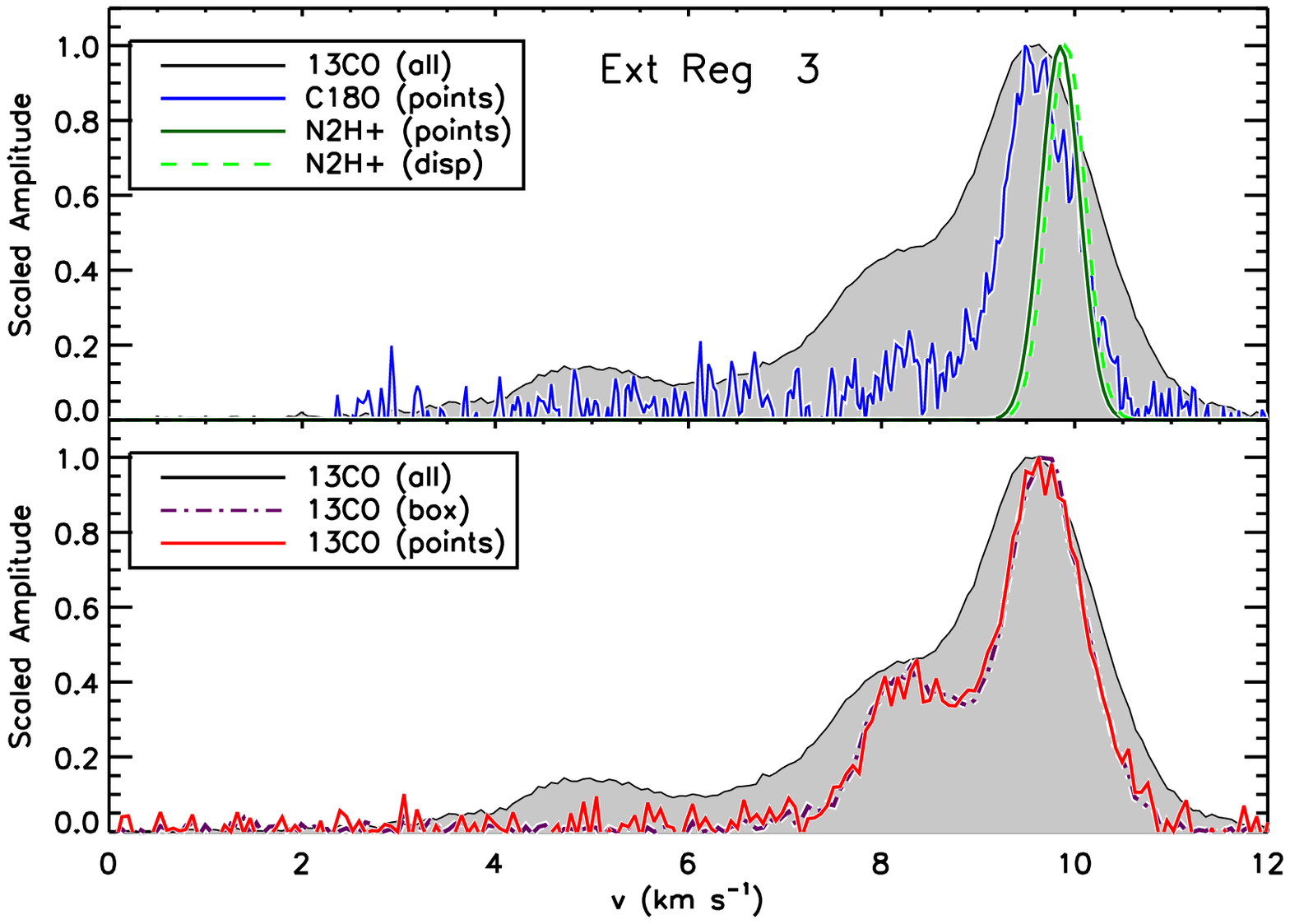}
\caption{Cumulative spectra within extinction regions 2 and 3.  See 
	Figure~\ref{fig_compare_spec_1} for the plotting conventions used.
	}
\label{fig_compare_spec_2}
\end{figure}

\begin{figure}[hp]
\plottwo{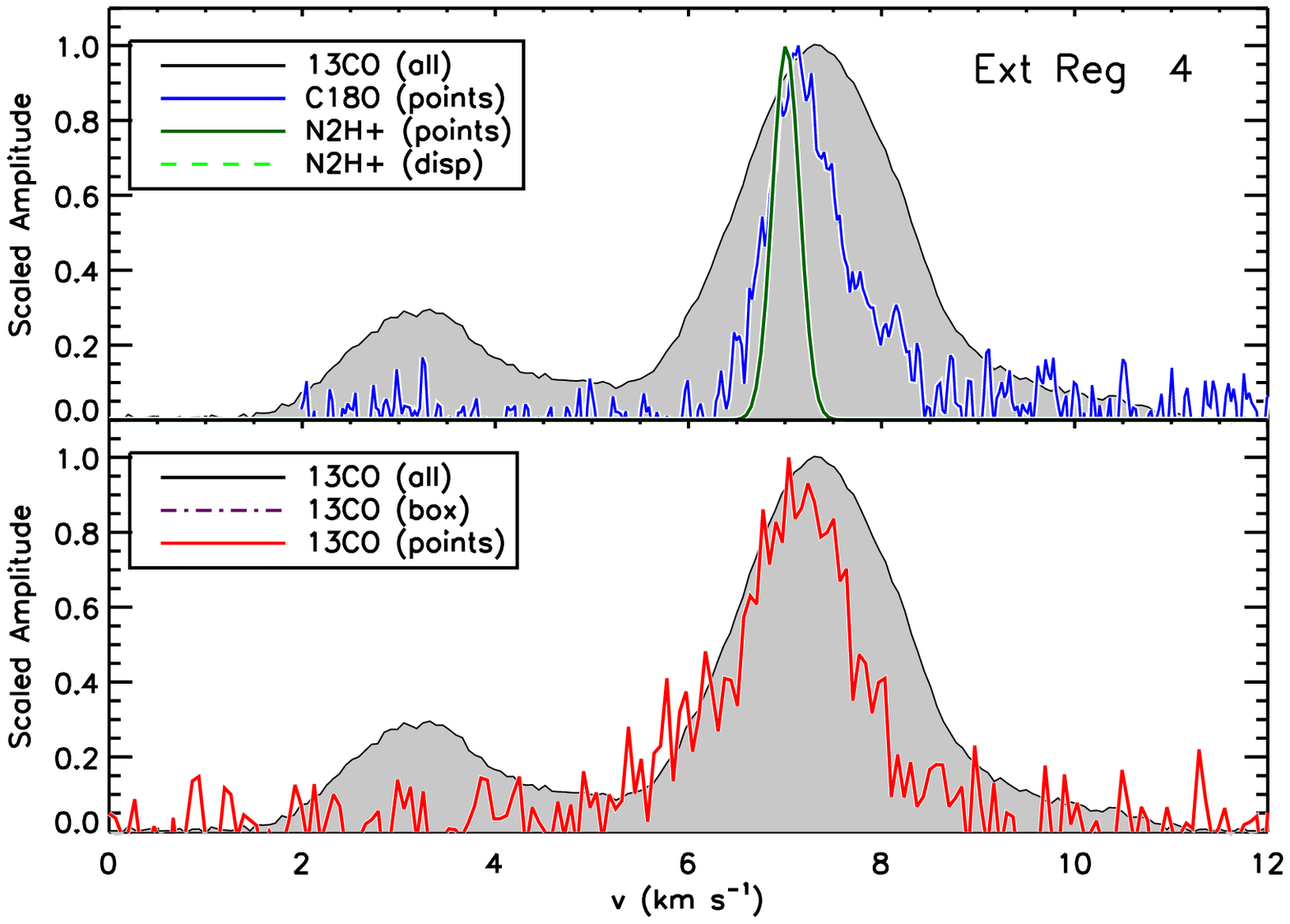}
	{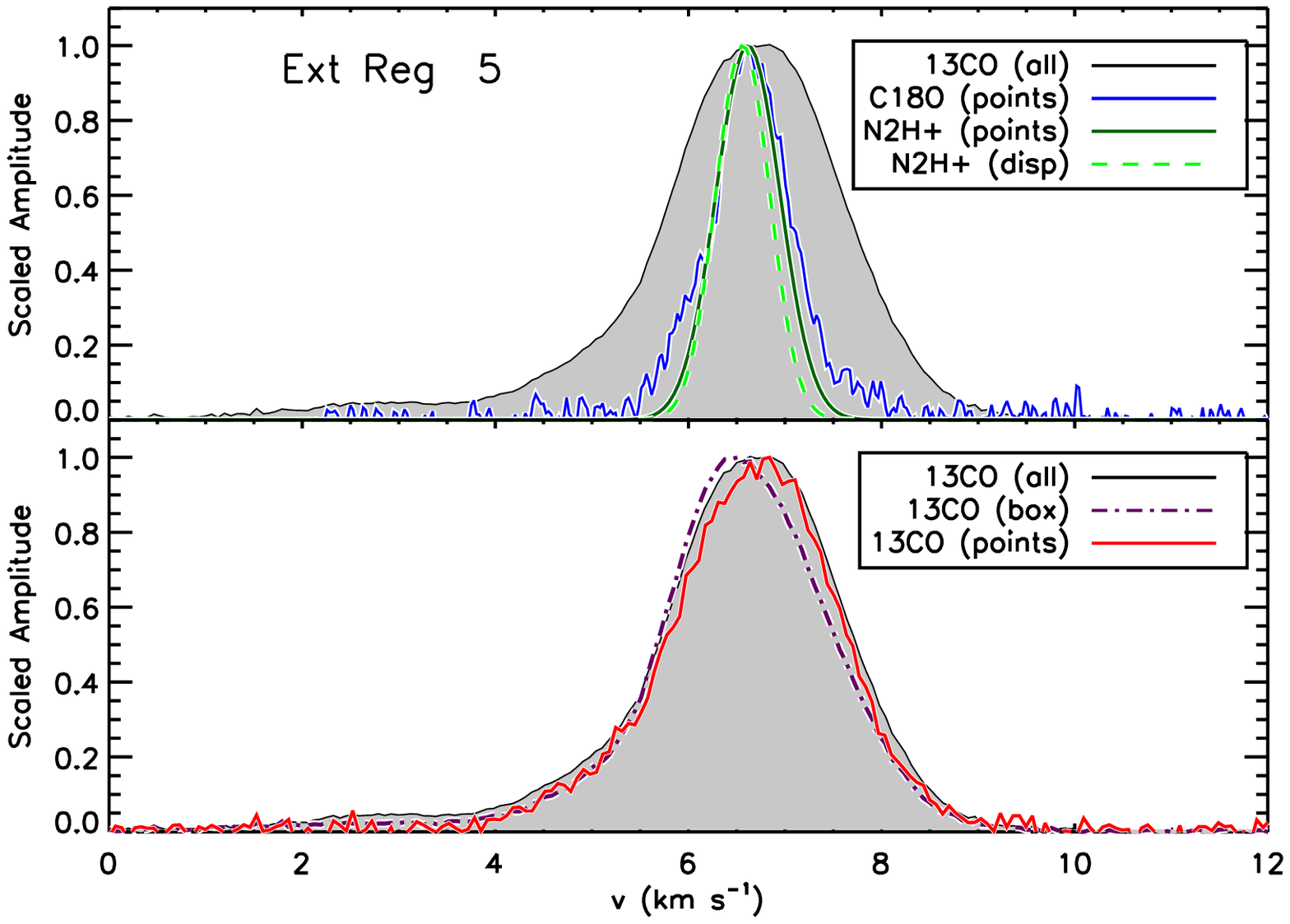}
\caption{Cumulative spectra within extinction regions 4 and 5.  See
	Figure~\ref{fig_compare_spec_1} for the plotting conventions used.
	Note that in extinction region 4, only one dense
	core was observed, so the standard deviation of \nh centroid
	velocities (dashed green line) cannot be plotted in the
	top panel.  The purple dashed and red solid line are
	identical in the bottom plot for the same reason.
	}
\label{fig_compare_spec_3}
\end{figure}

\begin{figure}[hp]
\plottwo{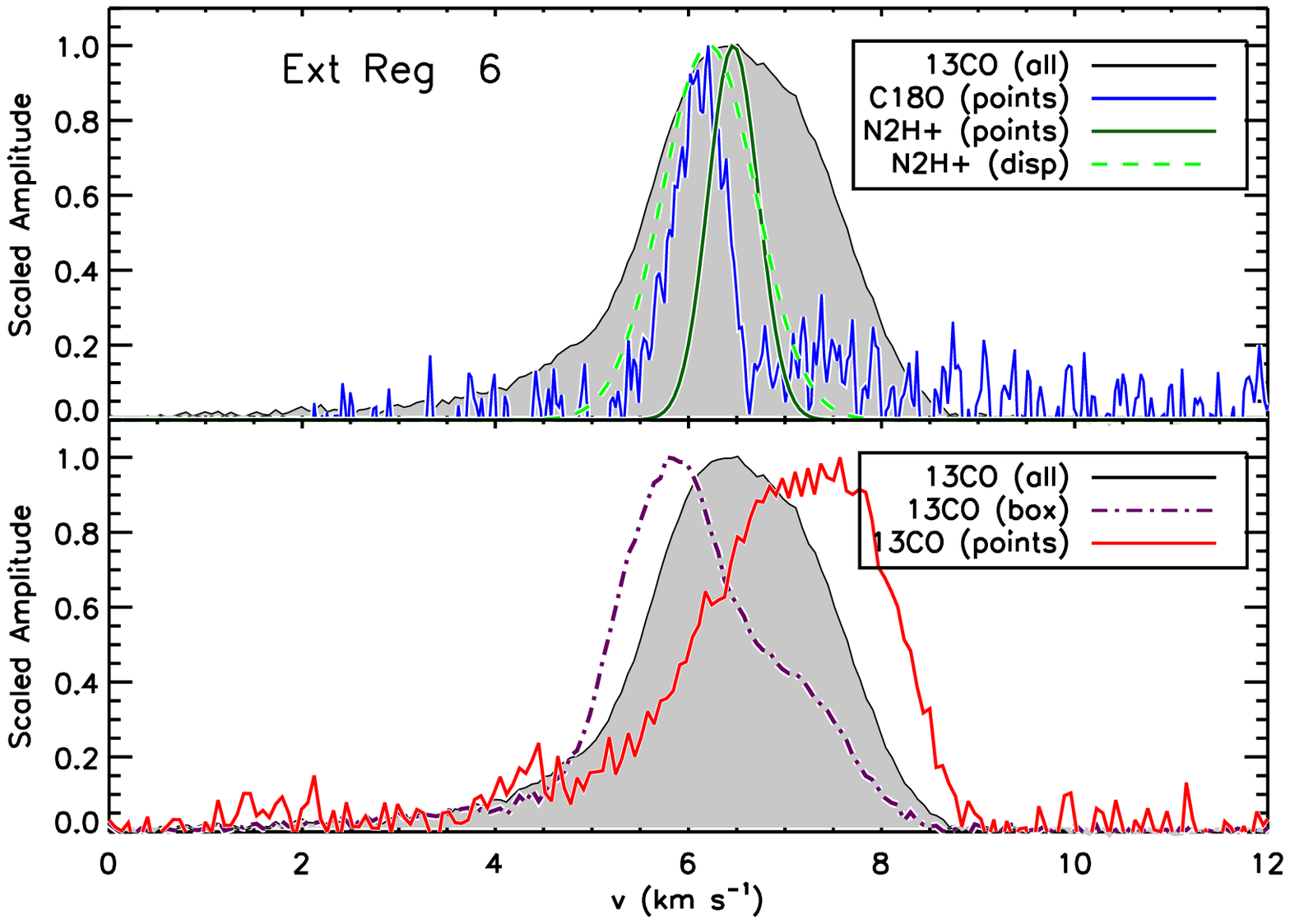}
	{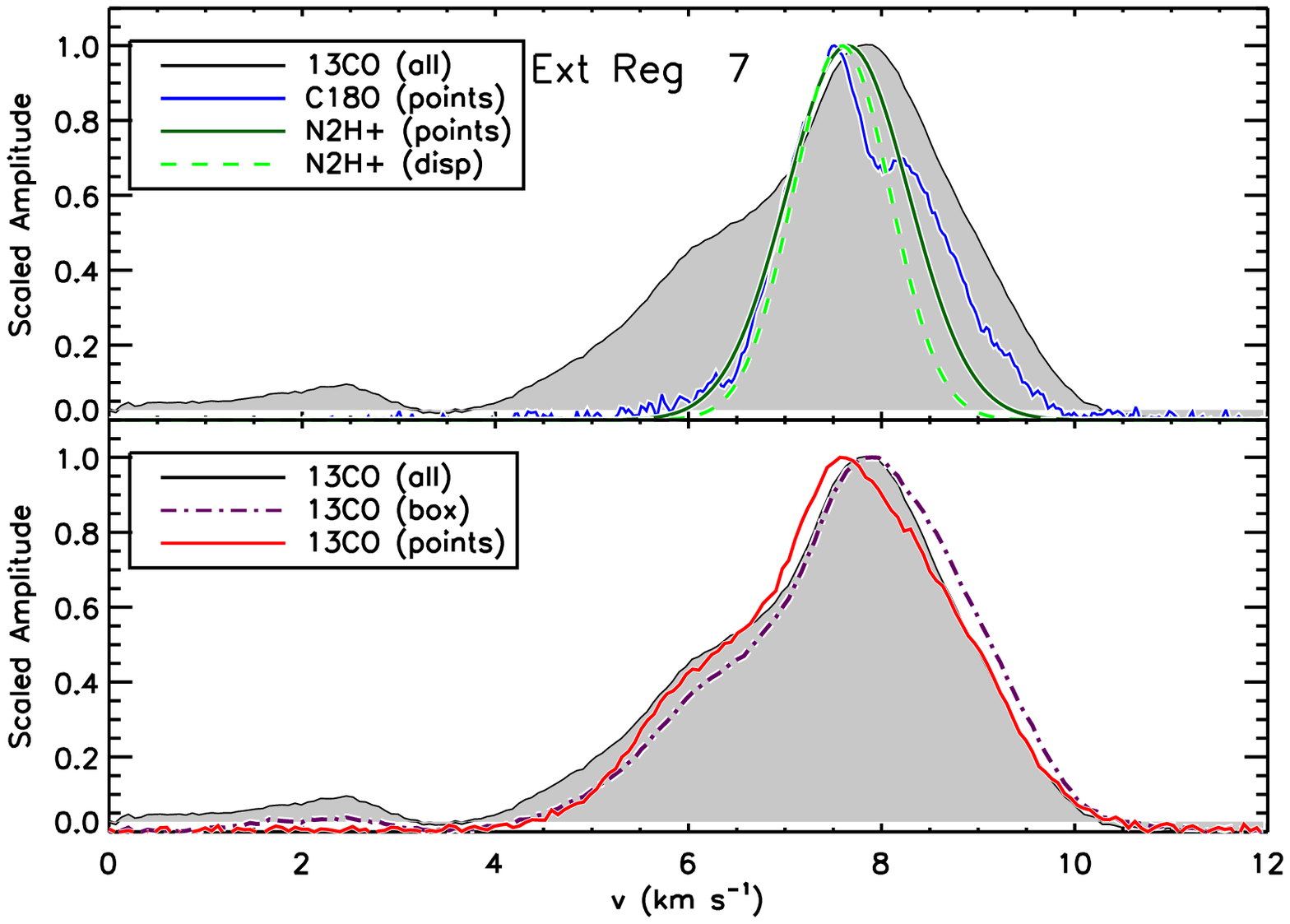}
\caption{Cumulative spectra within extinction regions 6 and 7.  See 
	Figure~\ref{fig_compare_spec_1} for the plotting conventions used.
	The offset in \nh centroids in extinction region 6
	is caused by the small number of cores (four) in the region,
	one of which is faint enough that it impacts the measurement
	using method A only.
	}
\label{fig_compare_spec_4}
\end{figure}

\begin{figure}[hp]
\plottwo{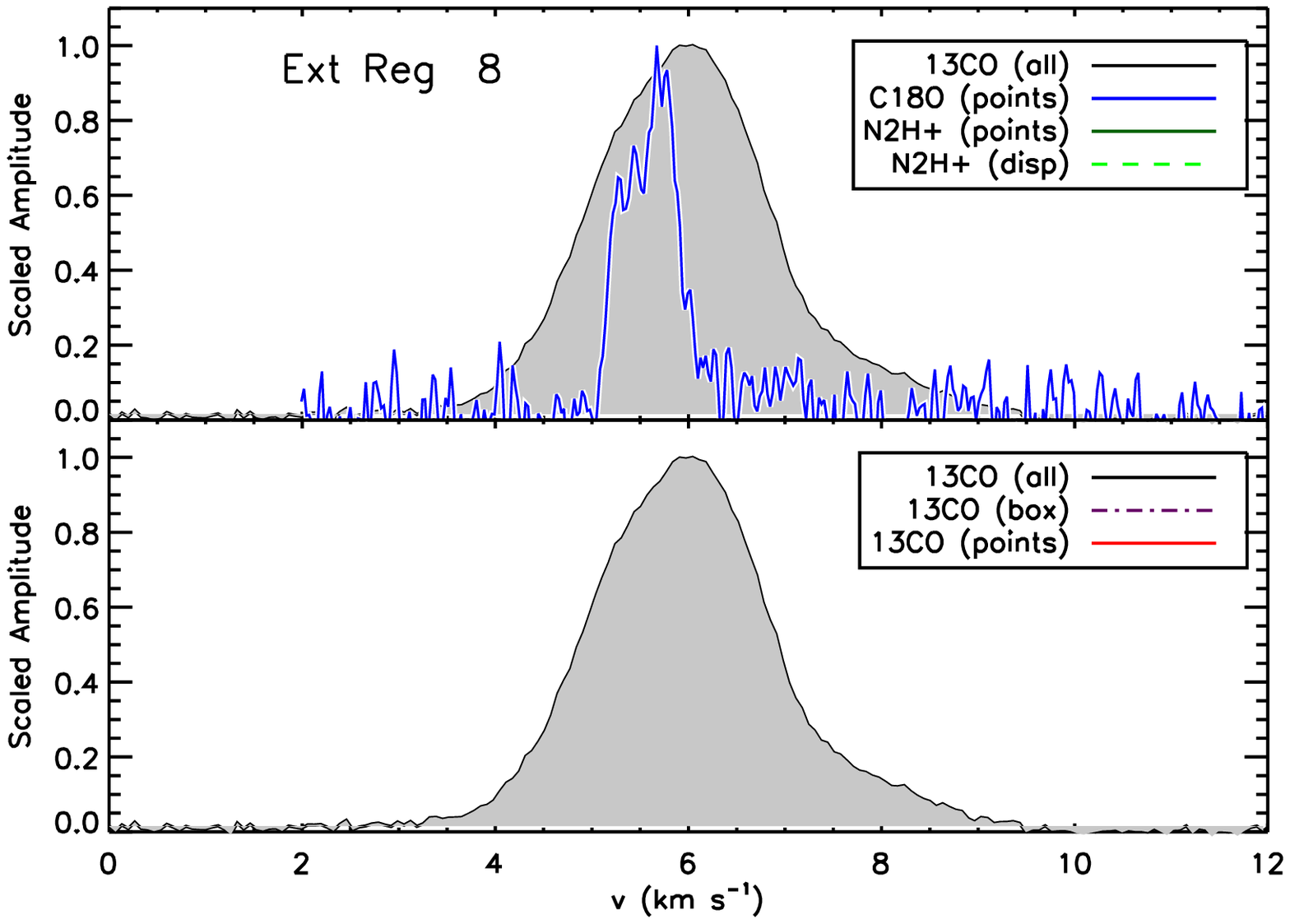}
	{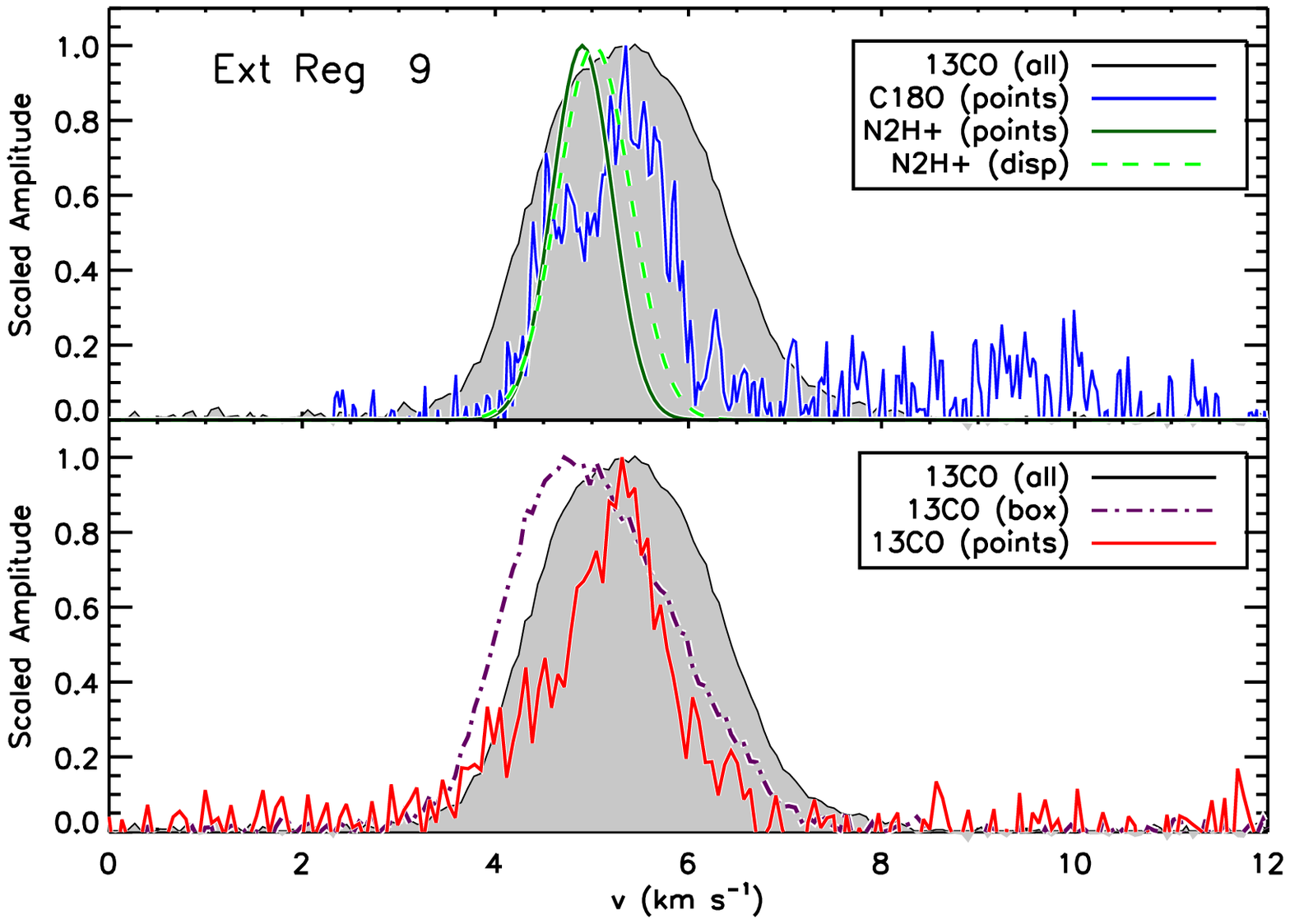}
\caption{Cumulative spectra within extinction regions 8 and 9.  See 
	Figure~\ref{fig_compare_spec_1} for the plotting conventions used.
	Note that there were no \nh detections in extinction
	region 8.}
\label{fig_compare_spec_5}
\end{figure}

\begin{figure}[hp]
\plottwo{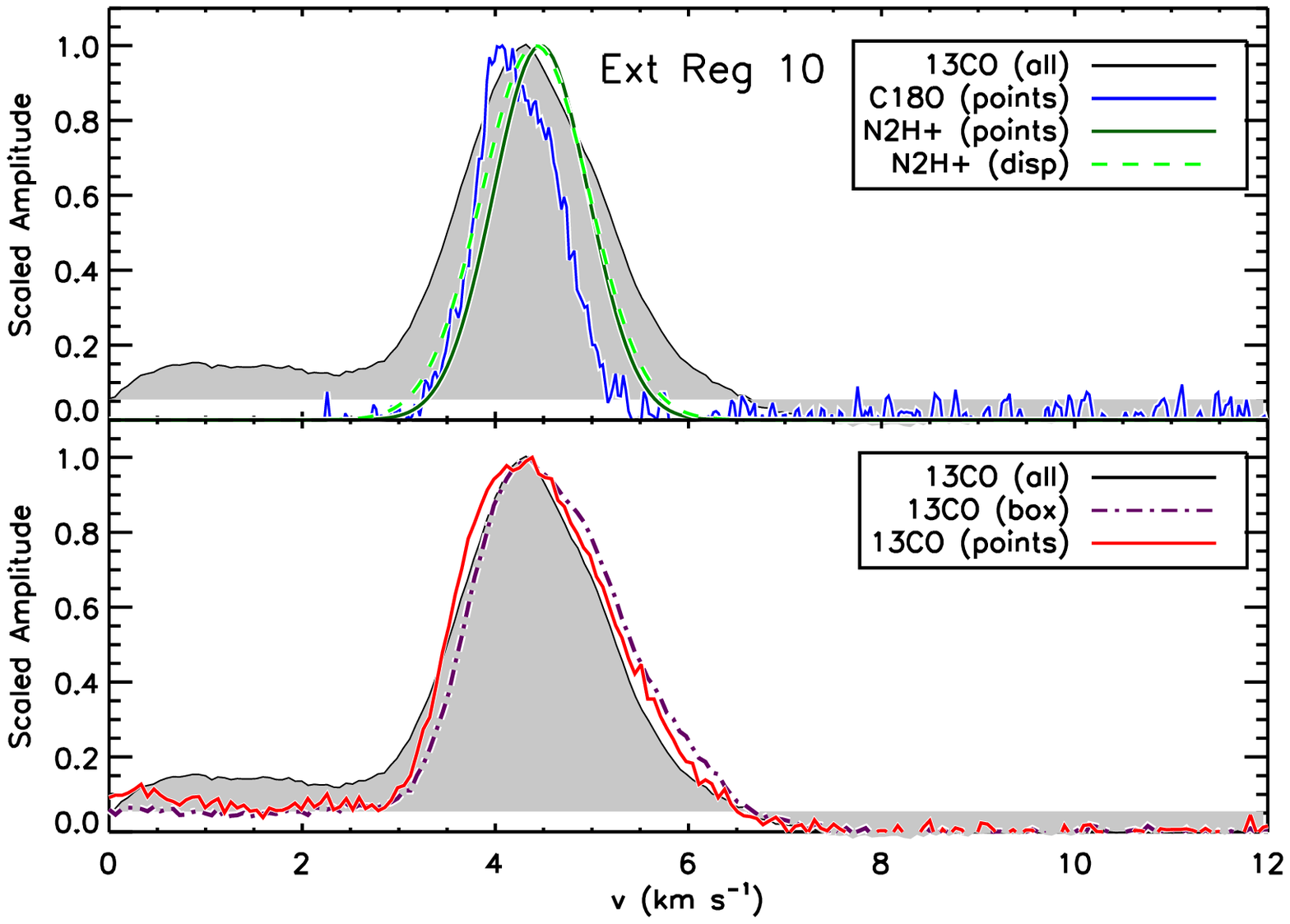}
	{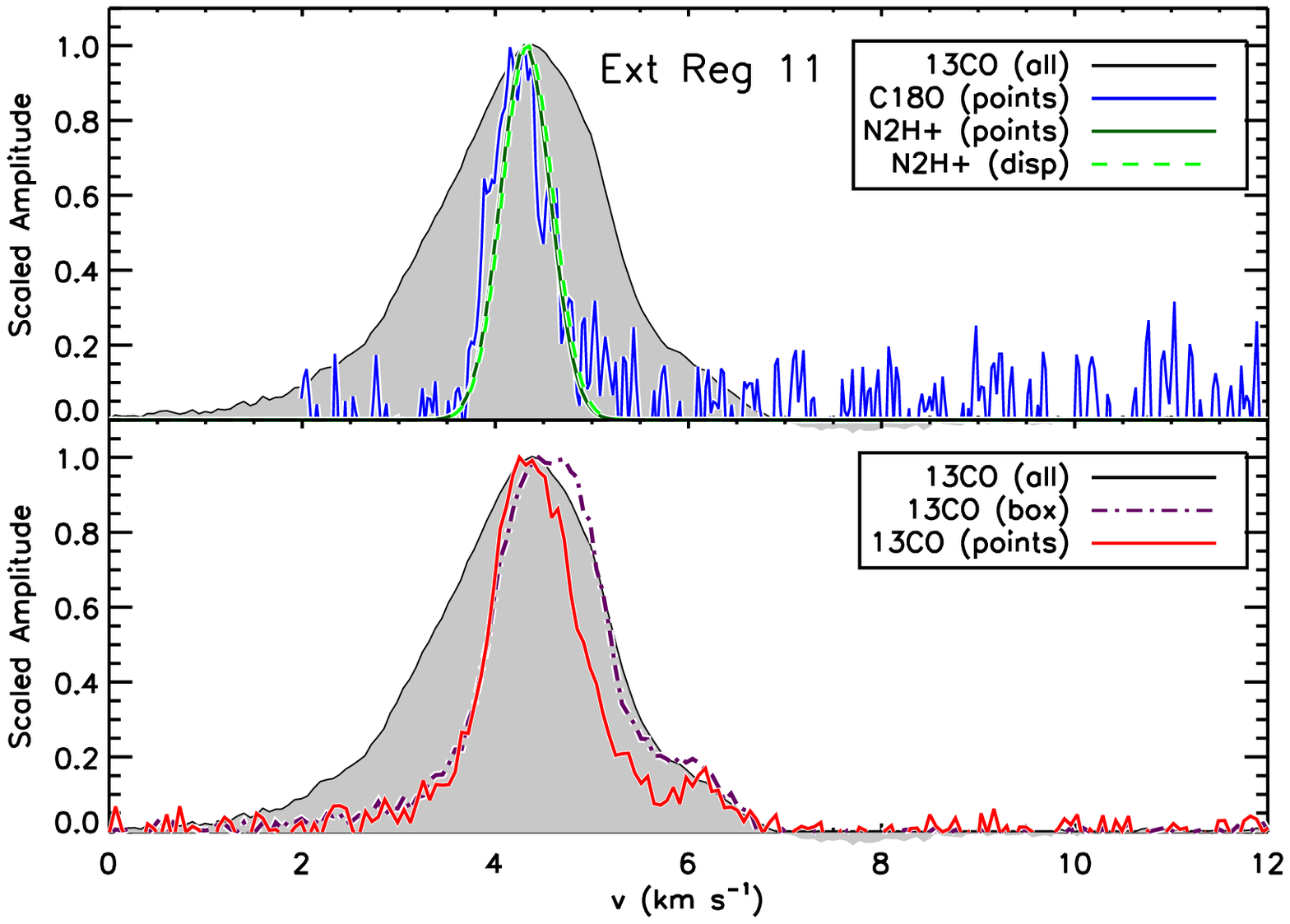}
\caption{Cumulative spectra within extinction regions 10 and 11.  See 
	Figure~\ref{fig_compare_spec_1} for the plotting conventions used.}
\label{fig_compare_spec_6}
\end{figure}

Figures~\ref{fig_compare_spec_1} through \ref{fig_compare_spec_6}
qualitatively show our main result (examined in more detail
below) -- the dense cores
have a much lower core-to-core velocity dispersion
than the velocity dispersion of the \thirco regions they reside in.
In the top panel of each figure,
a comparison is shown between the cumulative \thirco spectrum
across the entire region with measures of the dispersion seen
in all the \nh dense cores.  The dashed light green line shows the
result for \nh from method A above (the standard deviation of the 
\nh centroid velocities, centered on the mean centroid velocity), 
while the dark green line shows the result from method B above
(the dispersion of the sum of the \nh spectra).
As well, the dark blue line shows the result from
method B applied to the \co spectra in the region.
All of these cumulative spectra are compared to the region as 
a whole in \thirco (black and grey shading).
The figures show that the core-to-core velocity
dispersion is similar when measured using either method~A or
method~B, and that this velocity dispersion is significantly
smaller than the velocity dispersion of the region as a whole.

\begin{figure}[hp]
\plotone{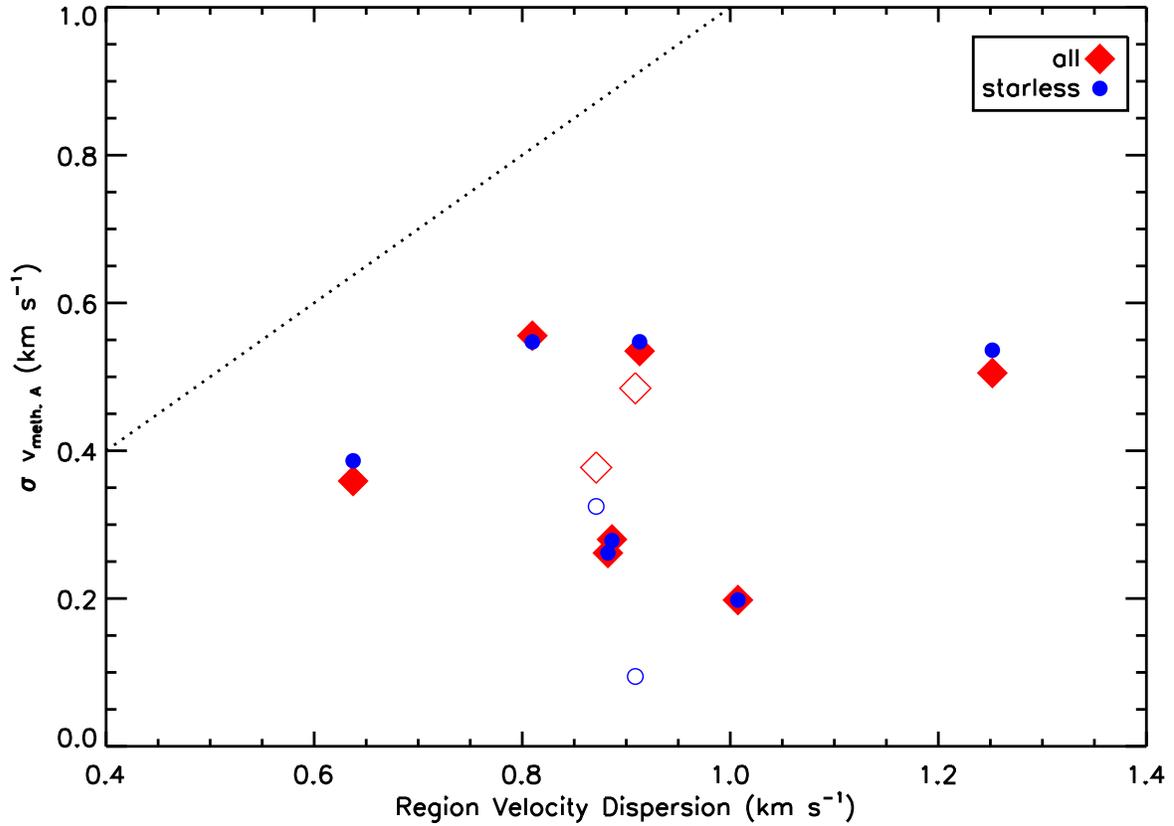}
\caption{The spread in core-to-core velocity dispersions
	found in each extinction region versus the velocity dispersion 
	of that region.  The diamonds show the values derived using
	all of the dense cores within each region, while the 
	circles
	show the values derived using only the starless cores within 
	each region.  The open symbols denote the values
	derived for the extinction regions with poor \thirco
	coverage.  The dotted line shows a 1-1 relationship.
	}
\label{fig_core_core_vs_region}
\end{figure}

We now determine the above result more rigorously.
In \citetalias{Kirk07}, we preferred the use of method A 
to measure the \nh core-to-core
velocity dispersion, due to the difficulties associated with method
B for \nh outlined in Section~5.1.  In that work, only starless cores were
considered, due to the potential
for protostars to have inherited more motion from surrounding
sources.  Here, we measure the standard deviation both for the
starless cores only as well as the full sample of cores (the full
sample was plotted in Figures~\ref{fig_compare_spec_1} to 
\ref{fig_compare_spec_6}).  Table~\ref{tab_core_core} provides
the values derived for the core-to-core velocity dispersion
for both method~A (both starless cores and the full sample)
and for method~B (full sample only).
Figure~\ref{fig_core_core_vs_region} shows a comparison of
the core-to-core velocity dispersion measured by method A
for the starless cores ($\sigma v_{meth. A,sl}$) 
and all cores ($\sigma v_{meth. A,all}$) versus the velocity
dispersion of the region ($\sigma_{13CO,reg}$). 
As can be seen in the figure, the core-to-core 
velocity dispersion tends to be similar in both samples;
variations between
the two may be attributed at least in part to small number statistics.
{\hknew The magnitude of variation expected due to small number
sampling is further discussed in the following subsection.}

{\hknew The standard deviation of core centroid velocities 
measured in each extinction region is a good description of the 
width of the distribution, despite the small number of cores 
typically present.  In the regions with a sufficient number of cores,
we tested this in two ways.  First, we
ran KS tests and found that the distribution of velocities in
each region was consistent with being drawn from a normal distribution
with a median and standard deviation equal to the value
measured, with over 80\% probability in most cases.  As a second test, we
compared the standard deviation to the median absolute deviation,
a robust measure of the width of distributions, i.e., little-effected
by outlying points \citep[e.g.,][]{Andersen08}.
For a normal distribution, the median absolute deviation is 1.48 times
smaller than the standard deviation; using this conversion factor,
we find the standard deviation as inferred from the median absolute
deviation agrees to within 12\%.  Based on these tests, it is
reasonable to assume that the standard deviation is a good descriptor
of the width of the core centroid velocity dispersion within each region.

As Figure~\ref{fig_core_core_vs_region} suggests, there is no
apparent variation with the core-to-core velocity dispersion 
with the region velocity dispersion.  In fact, all are consistent
with being originating from the same parent distribution.  Two-sided
KS tests of each pair of regions yields probabilities of 65\% in
all but one case (and 12\% in the remaining case) that the core
velocities are drawn from the same sample\footnote{All regions 
were compared with region
7, since that region has the largest number of cores.  The
median core velocity in each region was subtracted from all the core
velocities prior to comparison to account for the large-scale 
velocity gradient across Perseus.}.
}

\subsubsection{Effect of Sampling a Small Number of Cores}
\begin{figure}[hp]
\plotone{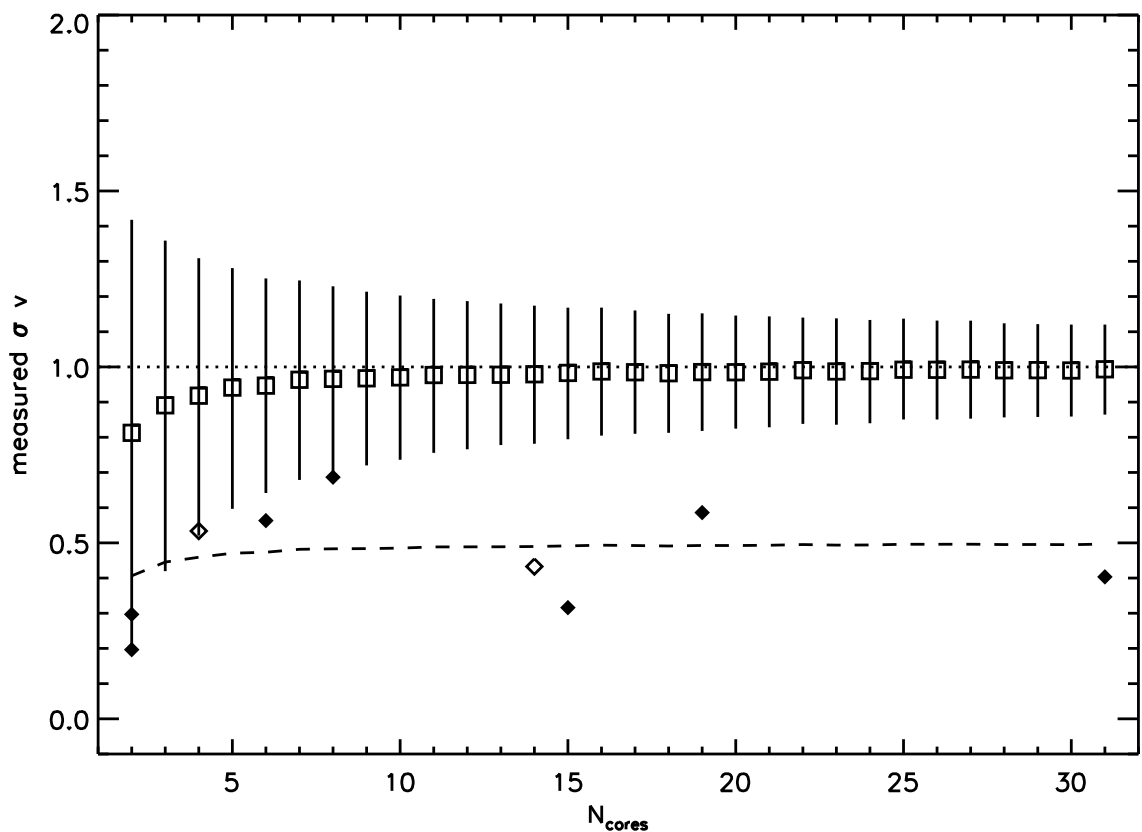}
\caption{The expected range of observed core-to-core velocity dispersions
	relative to the distribution they were drawn from for a given
	number of cores.  The boxes and vertical lines show the scaled
	mean and standard deviation of the core-to-core velocity dispersion
	for varying numbers of cores observed.  The diamonds show our
	observed values scaled to the velocity dispersion of the region,
	with the open symbols denoting the extinction regions with
	poor \thirco coverage.  The dotted horizontal line indicates
	an observed velocity dispersion equal to that of the parent
	sample, while the dashed line shows the scaled mean
	core-to-core velocity dispersion values multiplied by one half.}  
\label{fig_small_number_stats}
\end{figure}

The difference in size between the core-to-core and 
\thirco regional
velocity dispersion is sufficiently large that it cannot be 
attributed to the small number of cores within each region
{\hknew except for the regions with a very small number of cores.}
We ran a set of simulations to determine the size of the error
induced by using a small number of cores to measure the
underlying total velocity dispersion.
Scaling the regional velocity dispersion to a value of one,
we selected a sample of $N_{cores}$ objects with velocities
randomly drawn from a normal distribution (using IDL's
{\it randomn} function), and calculated the standard
deviation of the object-to-object velocity.
We ran 10000 simulations for each value of $N_{cores}$. 
Figure~\ref{fig_small_number_stats} shows the results of this
calculation.  The squares indicate the mean 
scaled velocity dispersion determined for 
the 10000 simulations for every value of $N_{cores}$, while
the vertical lines indicate the {\hknew 68.2\% confidence
levels, corresponding to the 1~$\sigma$ level in a normal distribution}.
Sampling a very small number of cores tends to increase
the difference between the measured and intrinsic velocity
dispersion, but does not systematically lower the measured
velocity dispersion by a significant amount when more than
{\hknew a few} cores are present, {\hknew in agreement
with the full analytic result from statistics \citep[see, e.g.,][]{Kenney39};
note that the full expression for the bias in the standard deviation
is more complex than the oft-quoted $\sqrt{n/(n-1)}$}\footnote{The
full correction factor is {\Large
$\sqrt{\frac{2}{n-1}} \quad \frac{\Gamma(\frac{n}{2})}{\Gamma(\frac{n-1}{2})}$} }
and is important to use for very low number samples.

The diamonds in Figure~\ref{fig_small_number_stats} show
the observed core-to-core velocity dispersions
(including both the starless cores and protostars), scaled 
by the \thirco velocity dispersion of the region. 
This figure clearly shows that all of the observed core-to-core
velocity dispersions lie well below the range of values
expected for a population of objects sharing the same velocity
dispersion as the parent \thirco gas.  
{\hknew Weighting the ratios between the core-to-core and region
velocity dispersions after correction
for the bias in the standard deviation gives a value of
0.5 $\pm$ 0.1, for either the full sample, or when restricted to
the regions with good spectral coverage.} 
This implies that if all regions have
similar ratios of core-to-core versus regional velocity 
dispersions, a ratio of around one half is a reasonable
estimate.

We conclude from this analysis that the dense cores tend to
have a much smaller core-to-core motion than is present in
the mean \thirco gas, roughly half the value, and that this
result is not due to small number statistics.

\subsubsection{Effect of Spatial Sampling}
Since the dense cores tend to be preferentially found concentrated 
within a small portion of the extinction region, it is reasonable
to ask whether the difference in size of the core-to-core and
total region velocity dispersions is caused by spatial sampling rather than
a physical difference between the two.  For example, the dense cores
could be at the centre of the potential well of the region and thus
naturally be expected to have a smaller velocity dispersion than
the mean gas spanning the entire potential well.  Here, we show that
the core-to-core velocity dispersions are intrinsically smaller than the
typical \thirco velocity dispersions, even accounting for the difference in
area sampled on the plane of the sky.

The bottom panel of Figures~\ref{fig_compare_spec_1} through 
\ref{fig_compare_spec_6} show the total \thirco spectrum
for each extinction region with differing areas considered.
The black line and grey shading (identical to the one in the bottom panel)
represents the sum of all \thirco spectra within
the region.  The purple line shows the result that would
be obtained if only the spectra within the area spanned
by the cores were considered.  In order to obtain this
estimate, we summed only the spectra that lie within the
minimal box (in RA and dec) 
containing all of the dense cores within the
region.  The resultant spectrum is similar to the total
(black) spectrum, and certainly is not similar in width to
the core-to-core result (green lines on the top panels).

We also considered an even more restrictive case -- the 
red line in the bottom panel of Figures~\ref{fig_compare_spec_1}
through \ref{fig_compare_spec_6} shows the resultant spectrum
if the \thirco spectra {\it only at the locations of the 
dense cores} are used in the summation.  This again
shows marked similarity to the spectrum for the region
as a whole, although with a lower signal to noise level.
{\hknew The \thirco spectra for each region when restricted
to the area spanned by or locations of the dense cores
can be somewhat narrower than the \thirco spectrum over the
entire region, however, this difference is not sufficient
to explain the smaller-still distribution of core-to-core
velocity dispersions within each region.  We measured the
ratio in the width (Gaussian sigma) fit to the areally-restricted
\thirco spectra versus the spectra for the entire region,
and find values of 88 $\pm$ 11\% and 86 $\pm$ 16\% for
the spectra for the region spanned by cores and at only the
core locations, respectively.  Since the core-to-core velocity
dispersion is approximately 50\% of the total region velocity
dispersion, the bulk of the difference cannot be explained
by the clustered locations of dense cores.}



\subsection{Interpretation}
Our analysis indicates that the core-to-core velocity dispersion
is significantly smaller than the velocity dispersion of the lower
density \thirco gas in the
larger extinction regions in which the cores are found.
Although this result is robust against the
potential biasses that we have investigated, interpretation
is hampered by a lack of knowledge of the full 3D structure
of the cloud.  If, for example, the \thirco within each
extinction region can be thought of as belonging to
a single, large, coherent entity, then the fact that the
core-to-core velocity dispersions are significantly smaller
than that of the entire region would indicate that the
dense cores are more bound than the region as a whole.  
On the other hand, if the \thirco gas actually comprises a
number of smaller, independent structures along and across
the LOS, then the \thirco may have a larger velocity dispersion 
than the cores because of the multiple structures probed along
the LOS.  
Even if the \thirco gas comprises a single entity along
the line of sight, if the dense cores probe only a fraction of
this material, a smaller velocity dispersion would be expected.
Larson's linewidth-size relationship \citep{Larson81}, 
for example, would predict a velocity dispersion half as large
for the cores as the region if the cores traced a length scale
roughly one quarter the length traced by \thircons.
We do not have any way to determine the
three dimensional cloud structure, hence the ratio
in core-to-core vs region velocity dispersion we measure cannot have a 
a definitive interpretation.  It is, however, a straightforward 
benchmark that can be used to evaluate star formation models.

\section{REGIONAL VELOCITY GRADIENTS}
\begin{figure}[p]
\begin{tabular}{cc}
\includegraphics[width=7.5cm]{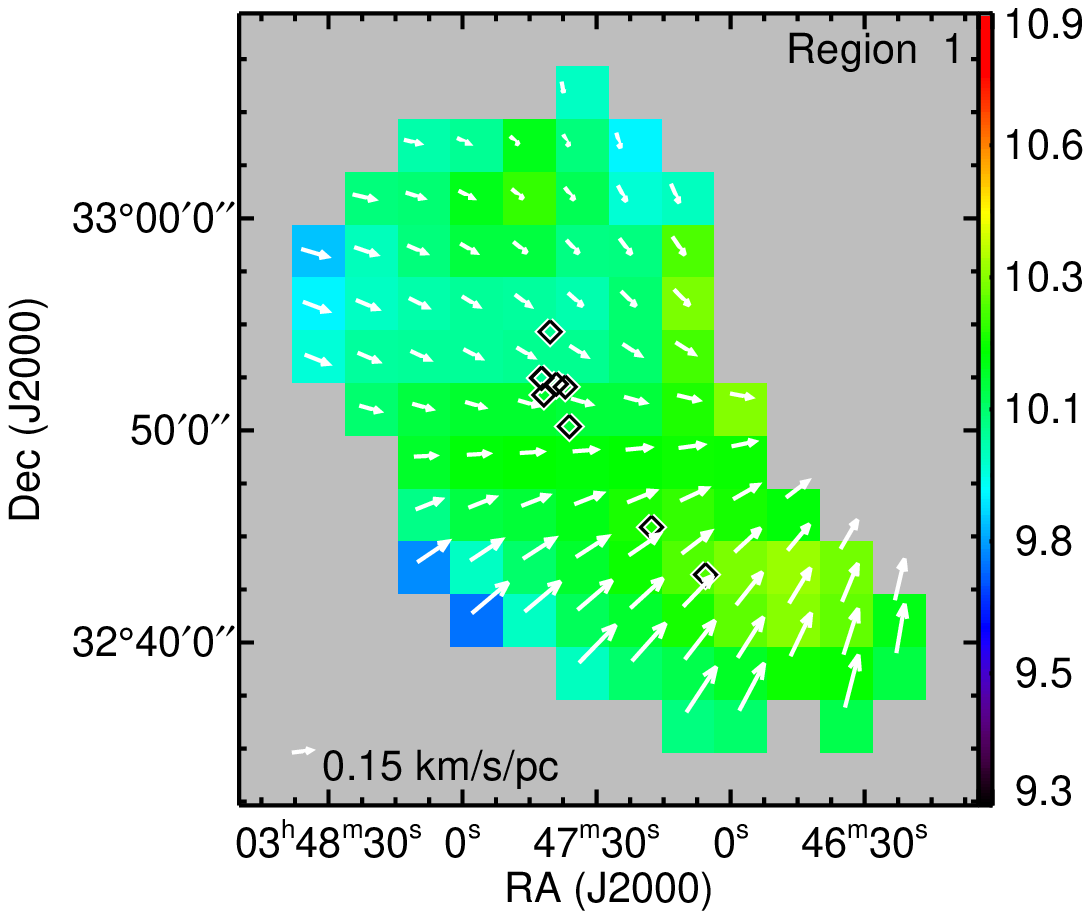} &
\includegraphics[width=7.5cm]{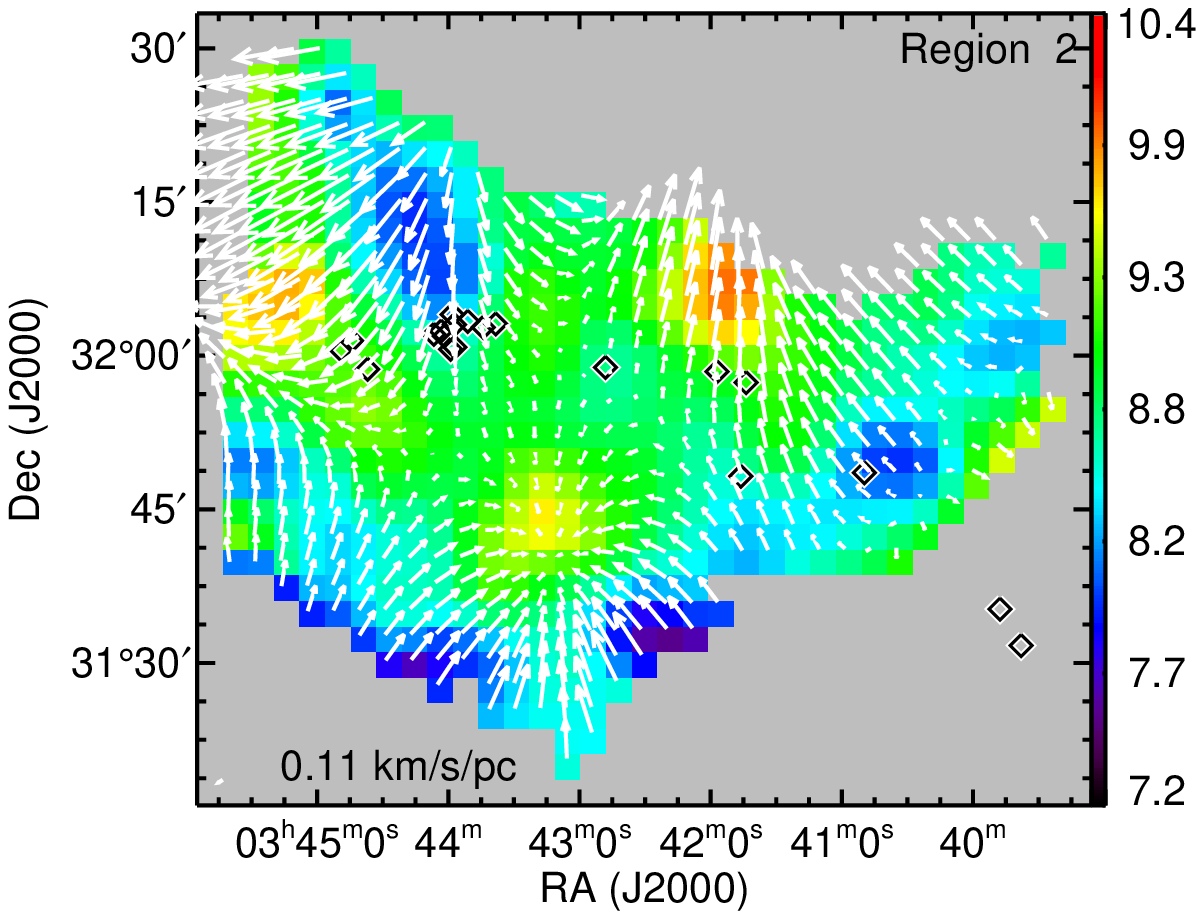} \\
\includegraphics[width=7.5cm]{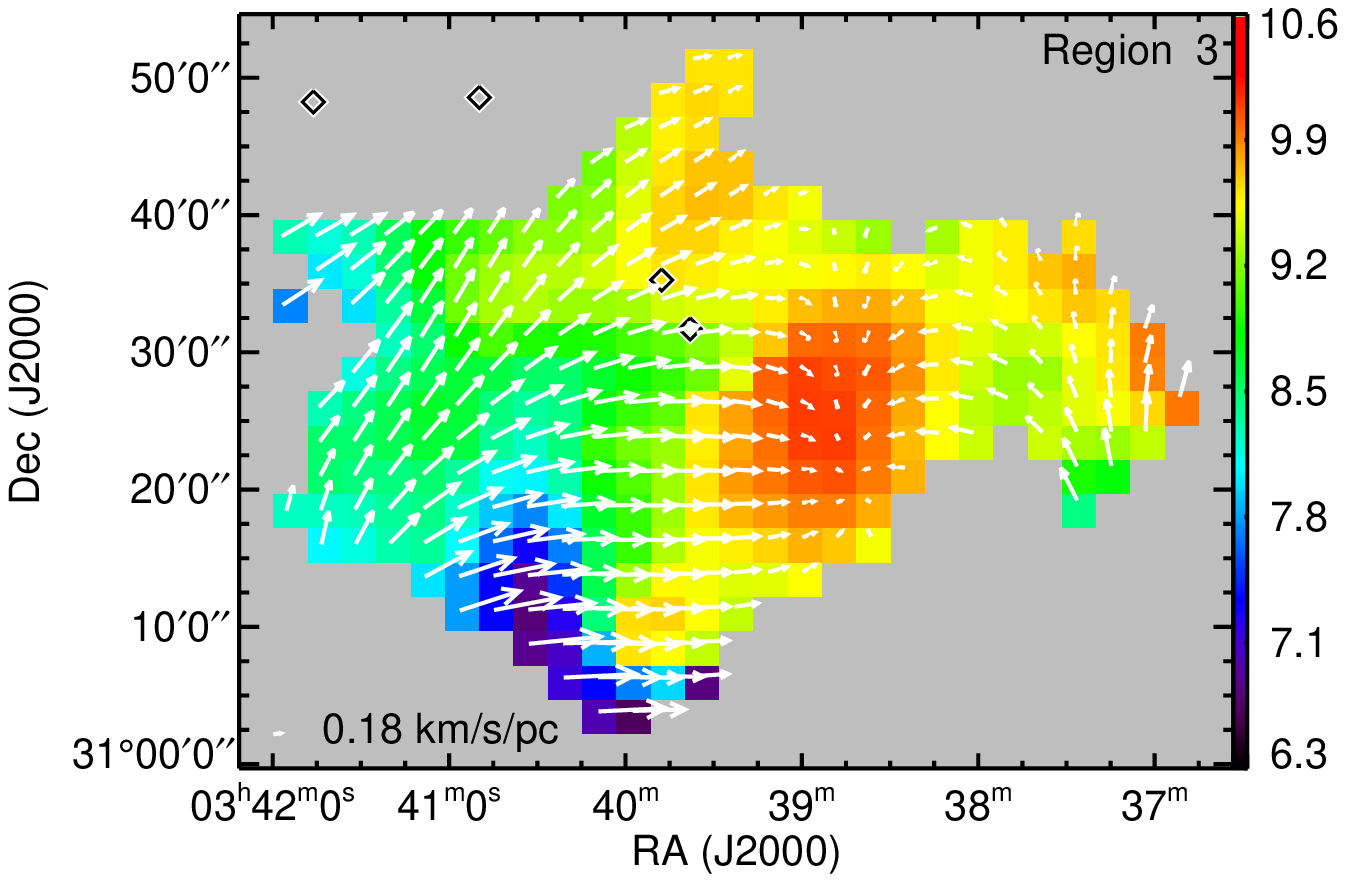} &
\includegraphics[width=7.5cm]{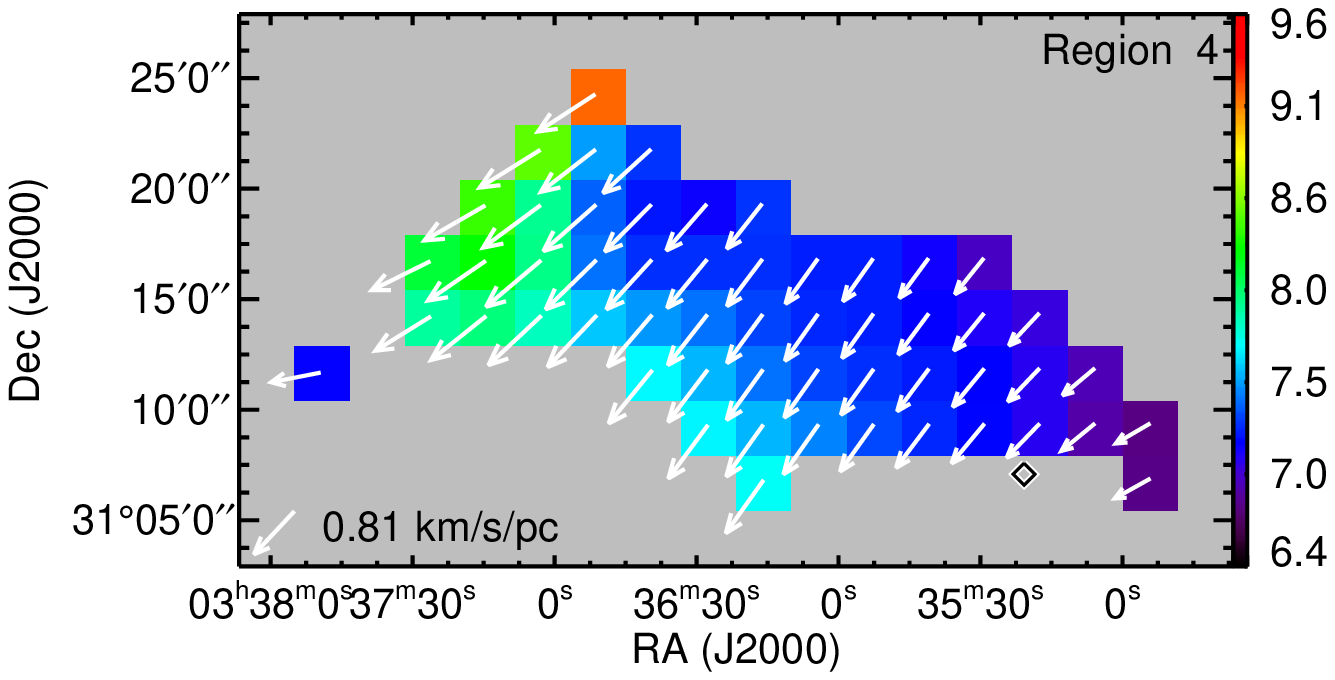} \\
\includegraphics[width=7.5cm]{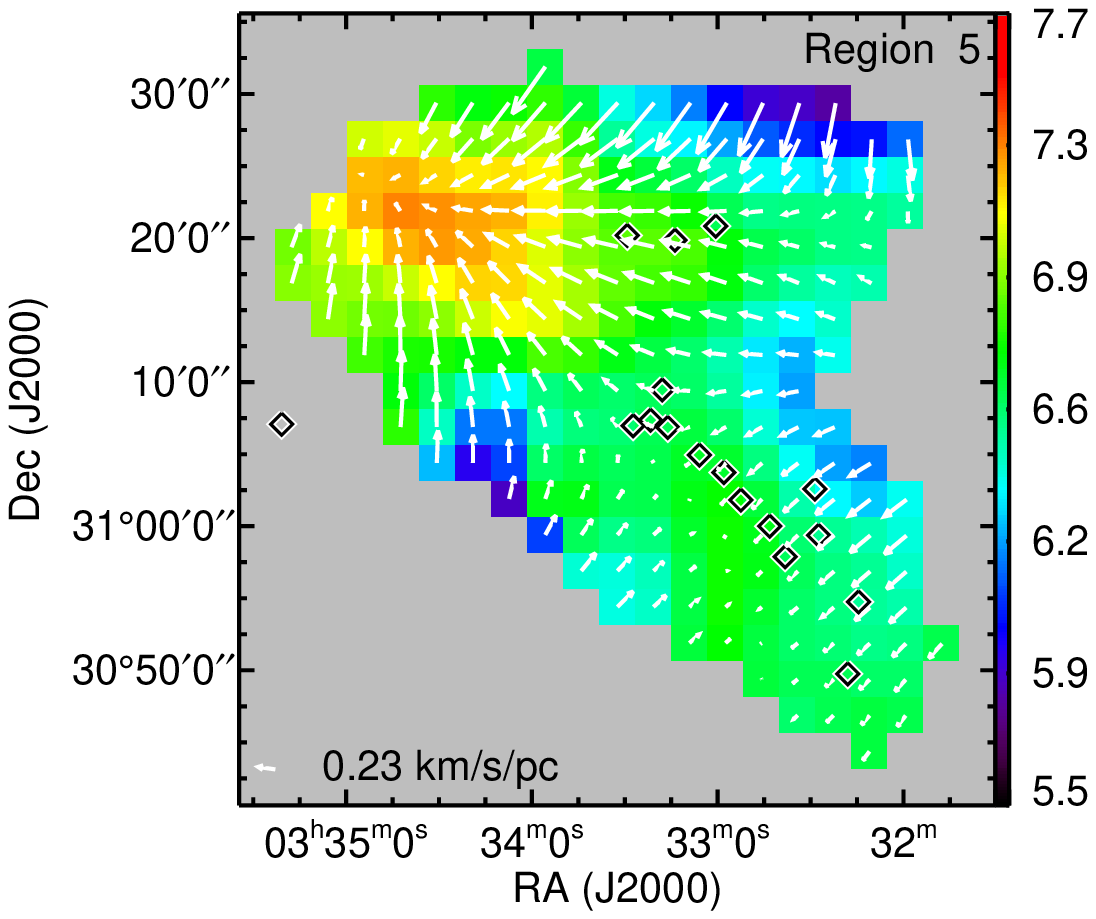} &
\includegraphics[width=7.5cm]{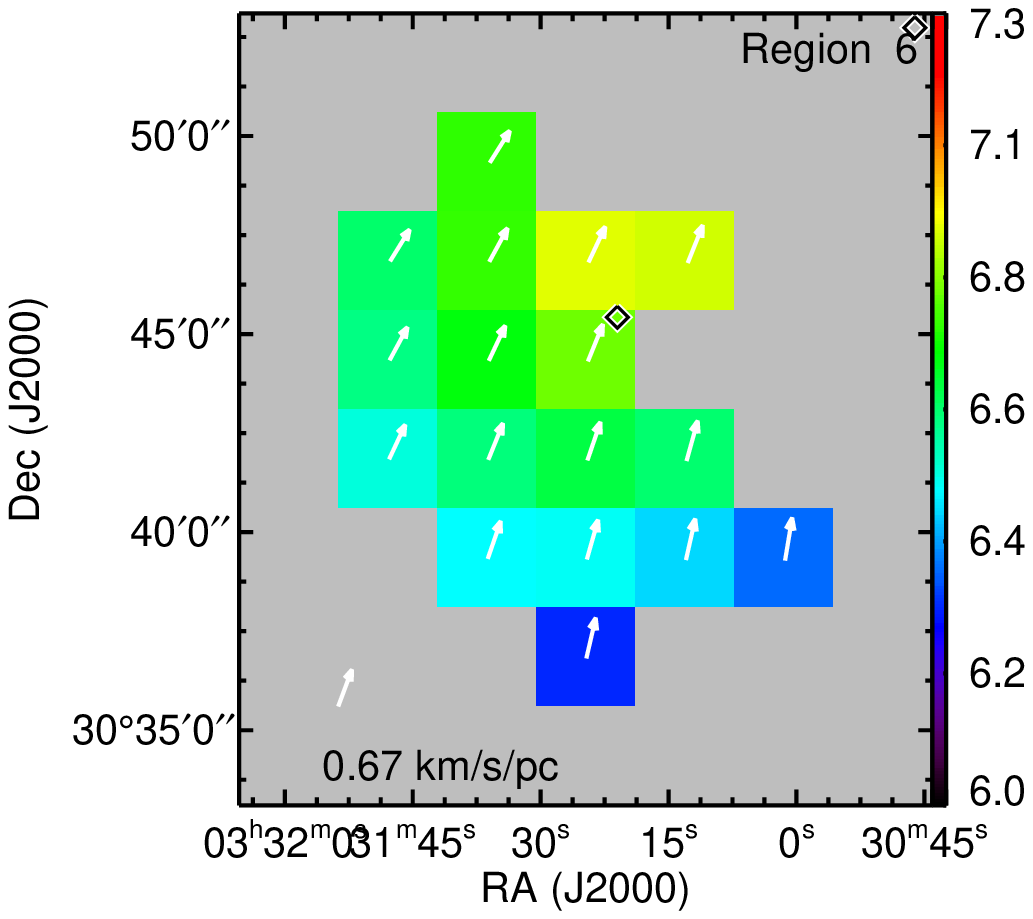} \\ 
\end{tabular}
\caption{The overall velocity gradient measured for each extinction region
	in \thirco for extinction regions 1-6.  
	The colourscale indicates the centroid velocity
	of each cell, while the arrows indicate the direction of the
	local gradient.  The overall gradient is indicated in the bottom
	left corner of each panel.  Diamonds show the positions of the
	\nh cores in each region.
	}
\label{fig_reg_grads_1}
\end{figure}

\begin{figure}[p]
\begin{tabular}{cc}
\includegraphics[width=7.5cm]{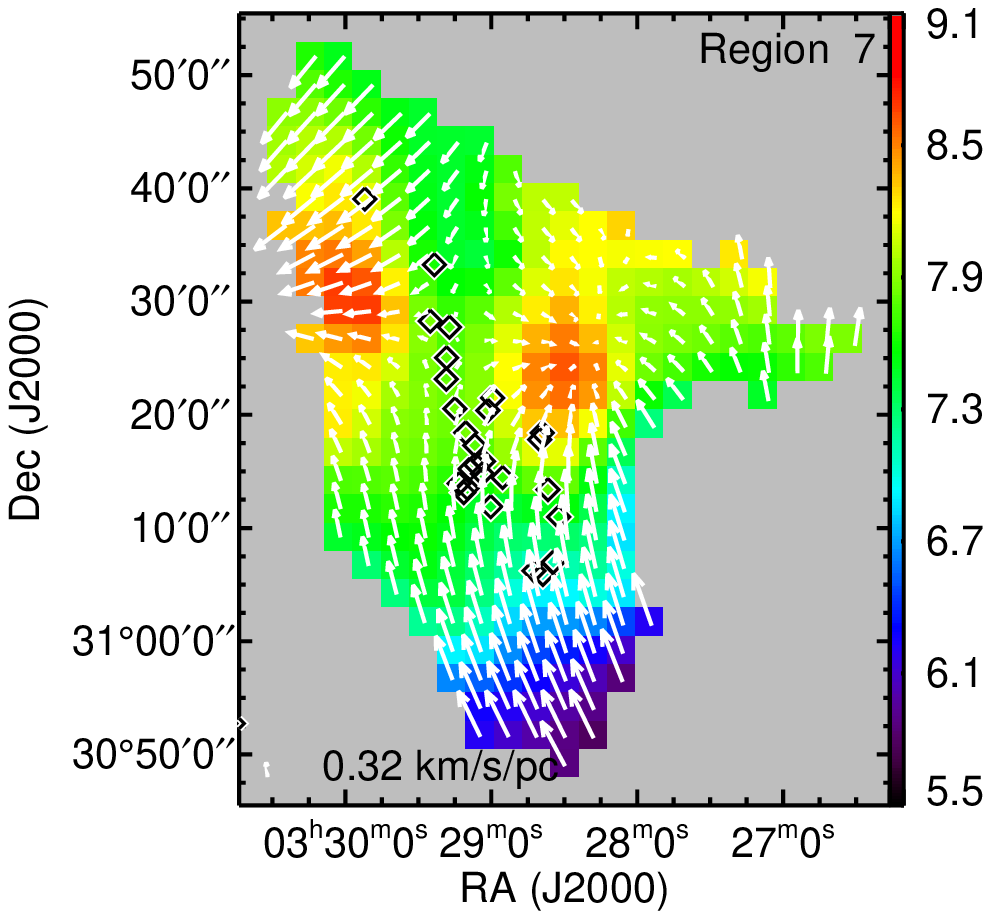} &
\includegraphics[width=7.5cm]{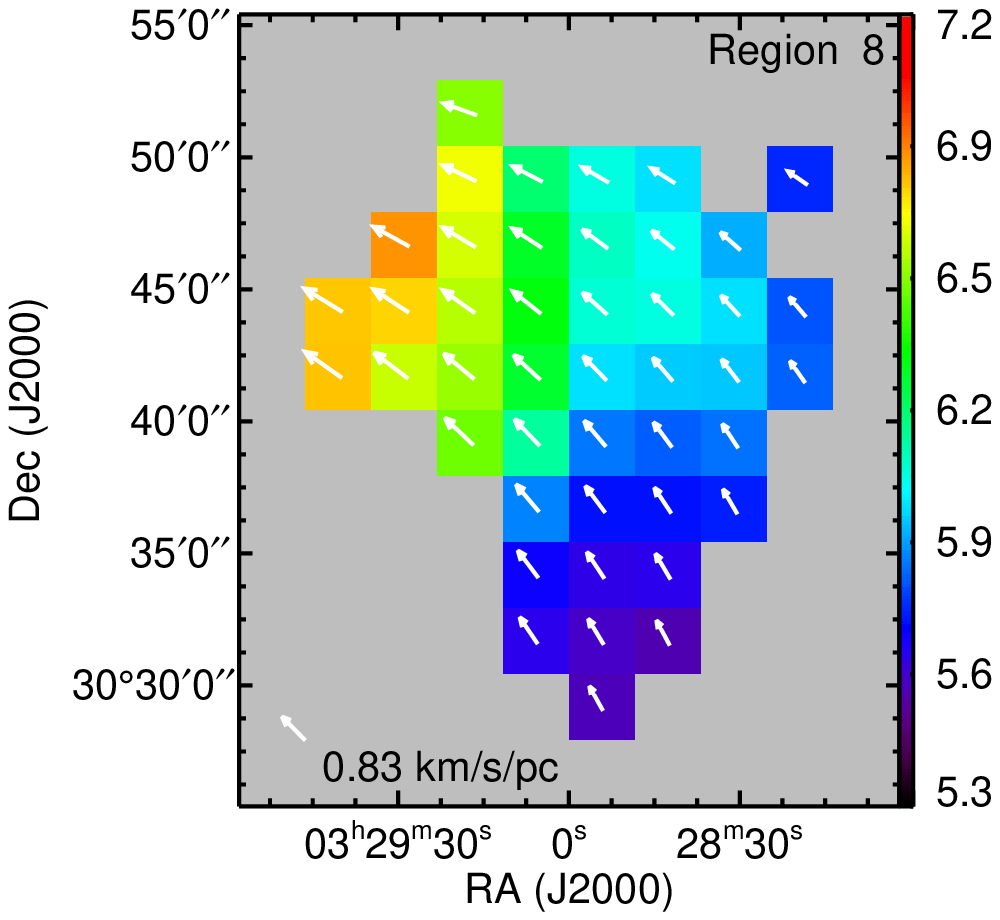} \\
\includegraphics[width=7.5cm]{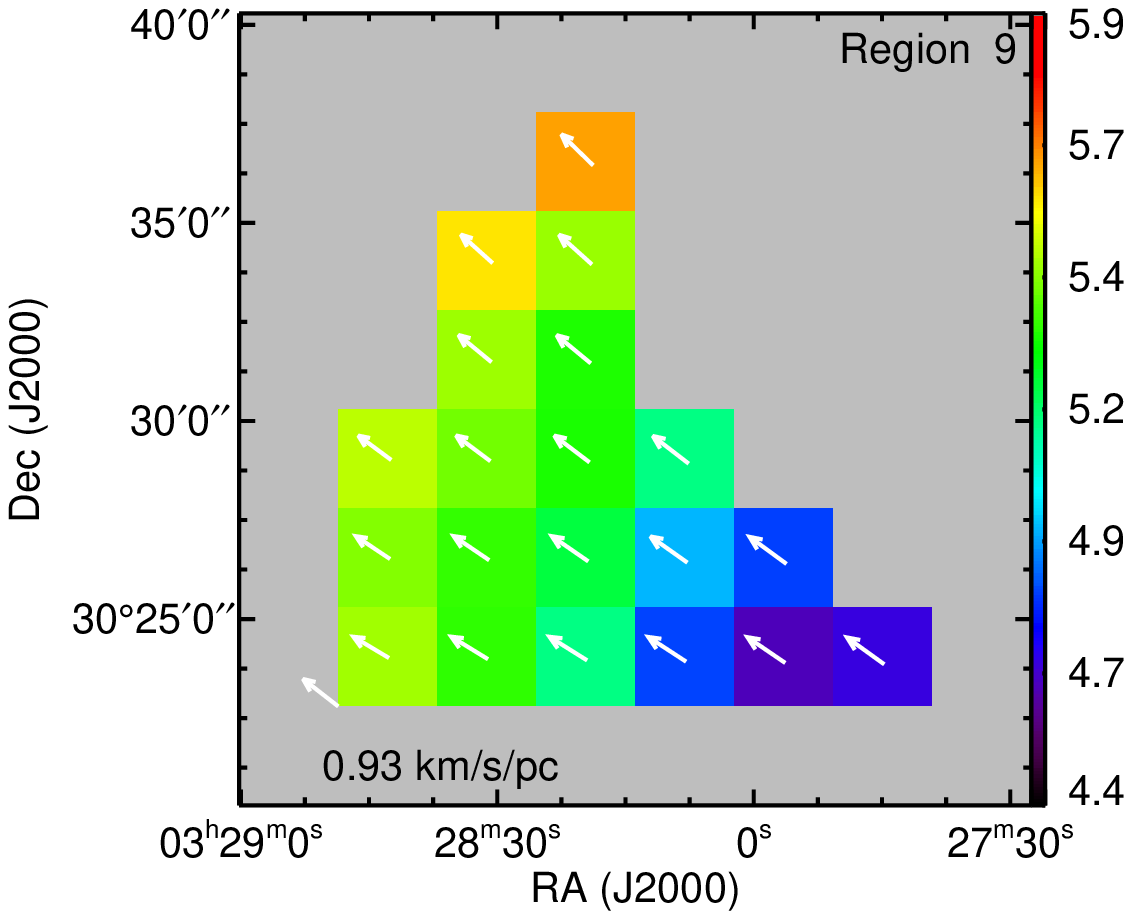} &
\includegraphics[width=7.5cm]{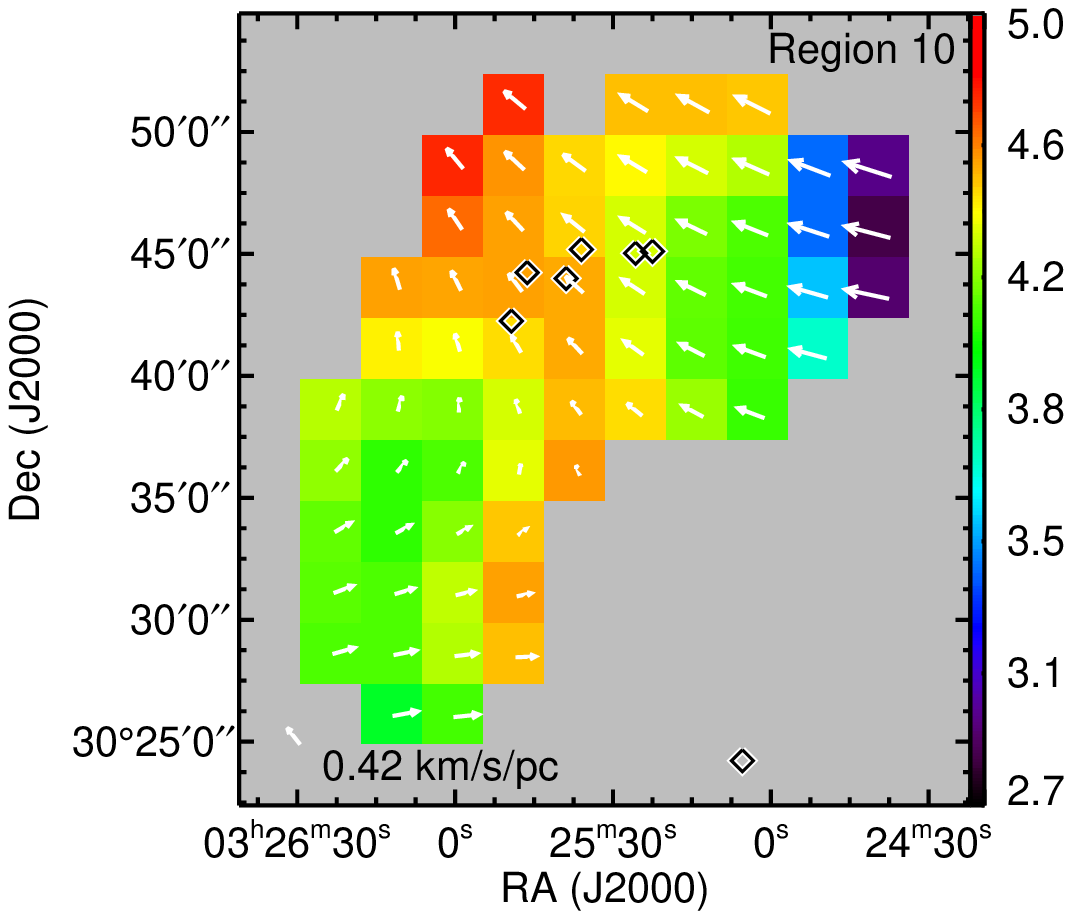} \\
\includegraphics[width=7.5cm]{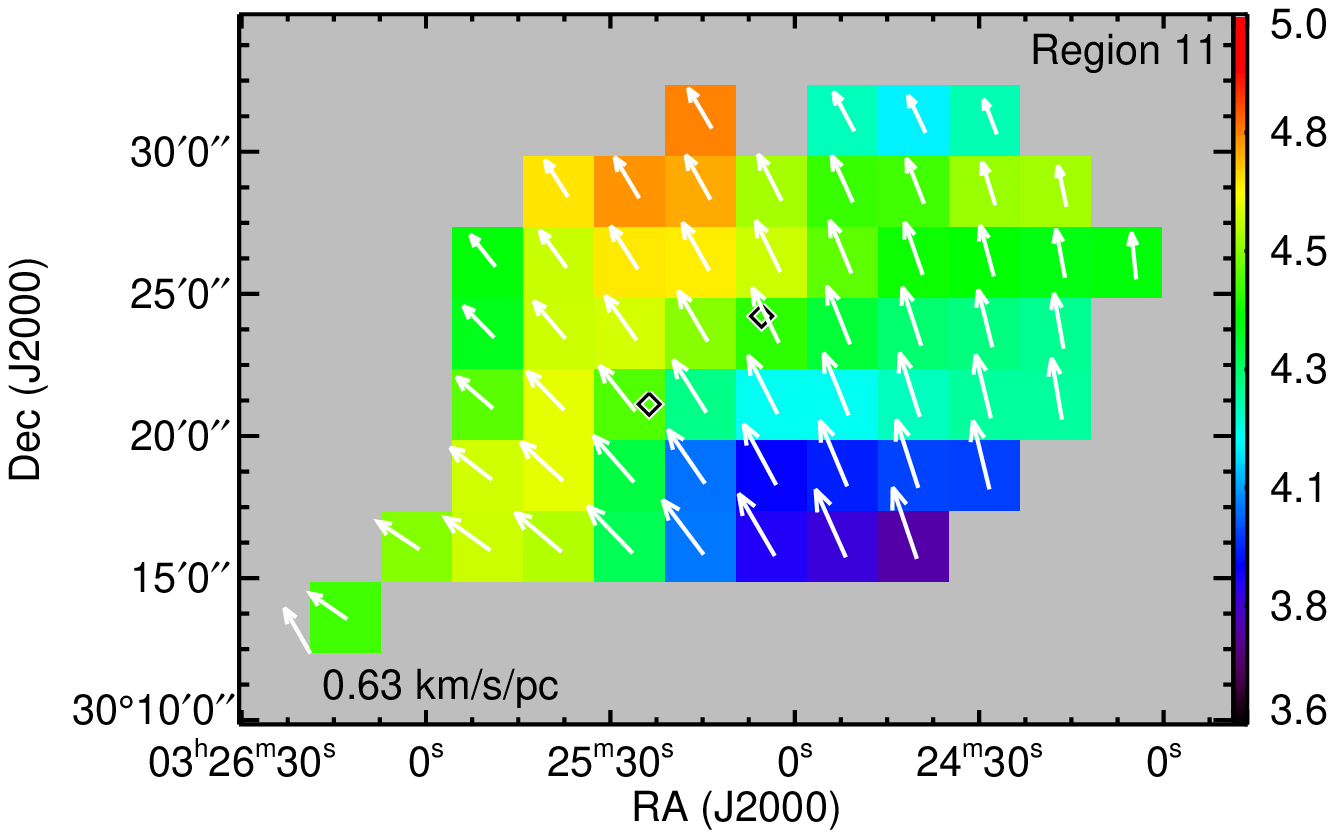} &
  \\
\end{tabular}
\caption{The overall velocity gradient measured for each extinction region
	in \thirco for extinction regions 7 to 11.  
	The colourscale indicates the centroid velocity
	of each cell, while the arrows indicate the direction of the
	local gradient.  The overall gradient is indicated in the bottom
	left corner of each panel.  Diamonds show the positions of the
	\nh cores in each region.}
\label{fig_reg_grads_2}
\end{figure}

Finally, we analyze the large-scale features in the velocity
structure of each extinction region.  
We calculate the
gradient of \thirco centroid velocities
across each extinction region using 
the 5\arcmin\ convolved \thirco data (see Section~4.1) and 
applying the gradient calculation method formulated in 
\citet{Goodman93}.  Figures~\ref{fig_reg_grads_1} and
\ref{fig_reg_grads_2} show the
centroid velocity measured for each cell in the convolved
\thirco map (colourscale) and the gradient measured
(white arrows).  The overall regional velocity gradient
is shown in the bottom left corner of each plot. 
Table~\ref{tab_tot_grads} provides the magnitude 
and orientation of the
gradient measured in each region, {\hknew along with the
formal error to the gradients fit; note that these errors are
often small due to the large number of points fit}.

\subsection{Cores Relative to the Gradient}
If the dense cores are connected to the large-scale
motions within each region, then they would be expected
to follow the same large-scale gradient as is observed
in \thircons.  We determine the velocity expected at
each position in the region by using the velocity at the
centre of the extinction region as a zero-point and extrapolating
outwards using the gradient measured, i.e.,
\begin{equation}
v_{flow,mod}(x,y) \quad = \quad v_{13CO}(x_c,y_c) + 
	\mathcal{G}  \cdot {\mathbf d}
\end{equation} 
or
\begin{equation}
v_{flow,mod}(x,y) \quad = \quad v_{13CO}(x_c,y_c) 
	\; + \; \mid \mathcal{G} \mid d_{\parallel}
\end{equation}

where $v_{13CO}$ is the velocity observed at the central
position $(x_c,y_c)$, $\mathcal{G}$ is the gradient 
(in km~s$^{-1}$~pc$^{-1}$), $d$ is
the distance from $(x,y)$ to $(x_c,y_c)$, and 
$d_{\parallel}$ is the distance from $(x,y)$ to 
$(x_c,y_c)$ in the direction parallel to the gradient's direction.

\begin{figure}[htb]
\includegraphics[height=15cm]{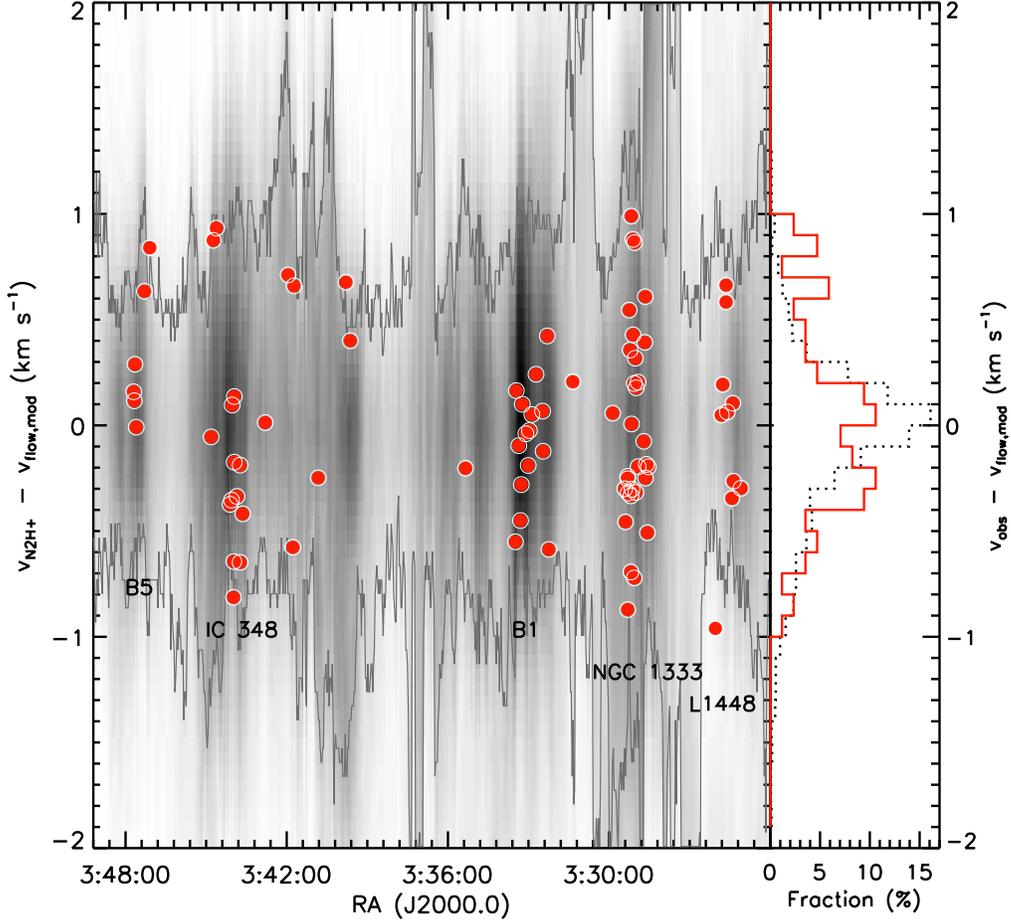}
\caption{ 
	The distribution of centroid velocity differences between
	the dense cores and the velocity inferred at their
	location from the \thirco velocity gradient measured
	in their extinction region (points) as a function of RA
	(left panel).
	The RAs of well-known star-forming regions are noted,
	with the exception of L1455 where the \thirco observations
	do not cover a large enough area to allow for a
	model of the gradient to be applied.  As in 
	Figure~\ref{fig_nh_vs_co_point}, the background greyscale
	image shows the PV diagram of the \thirco gas
	{\hknew and the grey contours show
        the half-maximum value at every RA}, with the
	spectrum at each RA shifted so that the peak emission
	falls at 0.  The total distrbution of $v_{N2H+} - v_{flow,mod}$
	is shown in the right panel in red, and for comparison,
	the total distribution of $v_{13CO} - v_{flow,mod}$ is
	shown in black.}
\label{fig_vals_vs_grads}
\end{figure}

Figure~\ref{fig_vals_vs_grads} shows the difference in
core centroid velocity, $v_{N2H+}$, and the velocity inferred 
at the core's position from the regional velocity gradient, 
$v_{flow,mod}$ as a function of RA.  As in Figure~\ref{fig_nh_vs_co_point},
the background greyscale image shows the PV diagram for \thircons,
with the value at each RA shifted so the peak intensity is at zero.
A similar scatter in the velocity difference is
found in all extinction regions.

The left panel of Figure~\ref{fig_vals_vs_grads} shows the 
distribution in velocity
difference between the cores and the regional flow,
$v_{N2H+} - v_{flow,mod}$ for the entire cloud (red), as well as 
the velocity difference between individual \thirco pointings and
the regional flow, $v_{13CO} - v_{flow,mod}$ (black). 
While the cores do not perfectly
follow the gradient-determined velocity, they tend to have a deviation
similar to that of the typical \thirco gas;
the standard deviations of both $v_{N2H+}-v_{flow,mod}$ and
$v_{13CO}-v_{flow,mod}$ are 0.45~km~s$^{-1}$.


\subsection{Large-scale Energetics}
The magnitude of the gradient across the region can be compared to the
total velocity dispersion measured for the region.  Regions
dominated by large-scale motions such as shear or rotation,
would be expected to have a high ratio of the velocity dispersion
inferred from the gradient to the total velocity dispersion, whereas regions
dominated by small-scale random motions would be expected
to have a lower ratio.

\begin{figure}[hp]
\plotone{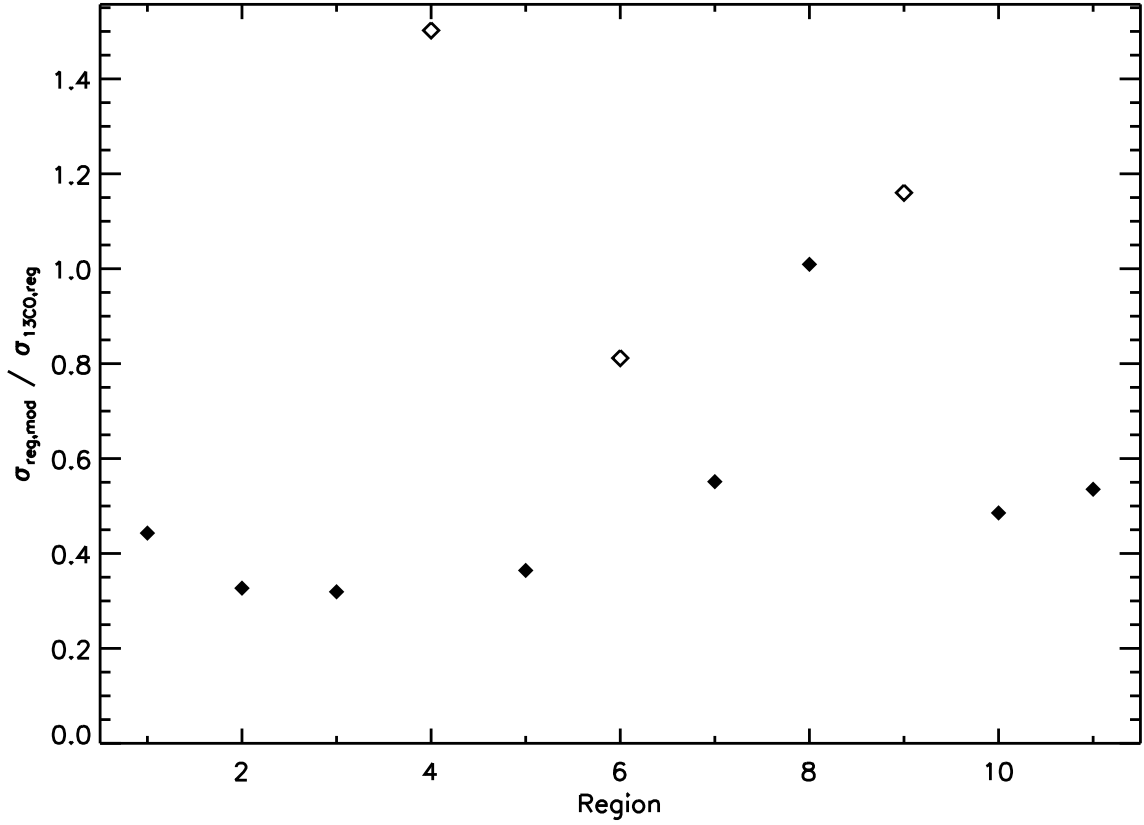}
\caption{The ratio of large-scale to total velocity dispersion observed in 
	each extinction region.  The open symbols show the
	results for the three extinction regions where the \thirco
	coverage is less than 80\% and hence less reliable.}
\label{fig_grad_vs_turb}
\end{figure}

{\hknew We approximate the velocity dispersion inferred from the gradient
across each region as the gradient multiplied by the diameter of the
region.  Since the \thirco emission across each region is, to first order,
constant, this is a (somewhat) better estimate than using, for example,
the radius of the region; all locations along the region contribute
approximately equally to the total region spectrum.  (Note that this
estimate would not be appropriate for centrally concentrated objects.)}
Figure~\ref{fig_grad_vs_turb} shows a comparison of the
velocity dispersion inferred from the gradient across 
each region 
versus the total dispersion observed in \thircons.
The ratio of
large-scale versus total velocity dispersion is around
50\% for the regions with well-determined measurements.
A few of the regions show ratios above 100\%, which can
be understood by the fact that
our estimate of the large-scale velocity
dispersion is crude.  
For example, a varying column density across
a region would tend to weight a limited range in velocities,
and hence decrease the total velocity dispersion measured,
while leaving the large-scale velocity dispersion determined
from the gradient unchanged.


Turbulent motions that are dominated by large-scale modes
can also produce an observable gradient. 
\citet[][hereafter BB00]{Burkert00}
investigated the magnitude of the gradient expected from
large-scale turbulent modes for cores similar
to the NH$_3$ cores observed in \citet[][hereafter G93]{Goodman93}, 
\citet[][hereafter BG98]{Barranco98}, and \citet{Goodman98}.
BB00 ran several
thousand realizations of a 3D turbulent velocity field with 
power spectra of 
$P(k) \propto k^{-4}$ and $P(k) \propto k^{-3}$,
where $P(k)$ is the power at wavenumber $k$; the oft-adopted
Kolmogorov (incompressible turbulence) power spectrum falls
within this range with  
$P(k) \propto k^{-11/3}$.
Along with this turbulent velocity field, BB00 adopted a spherically
symmetric, centrally-condensed density distribution.
They then calculated the resultant
spectra observable in 2D, and determined the gradient across the
core using the 
same method as described in \citetalias{Goodman93} (and also used in
our analysis).
The $P(k) \propto k^{-4}$ turbulent power spectrum has more
power in the largest modes, and, as expected, was shown
to generally produce larger observed gradients.

\begin{figure}[htb]
\plotone{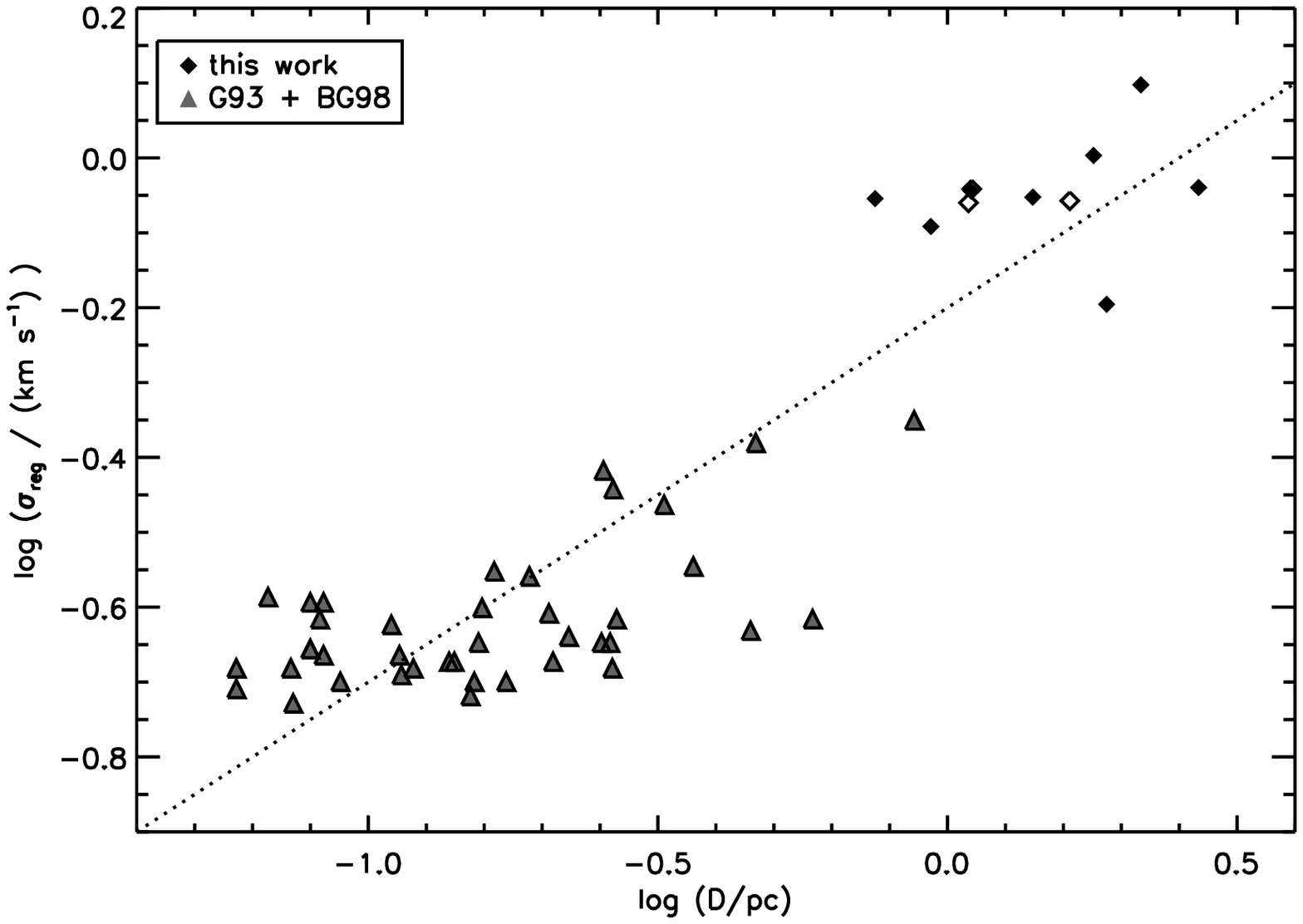}
\caption{The observed velocity dispersion versus diameter for all of the
	extinction regions (diamonds) and NH$_3$ cores in \citet{Goodman93}
	(triangles).  The open symbols show the results for the regions
	where the \thirco coverage was less than 80\%.  The dotted
	line shows the slope of the scaling relationship 
	assumed by \citet{Burkert00}, $\sigma \propto R^{0.5}$.
	Note that a scale factor for the relationship is not given in 
	\citet{Burkert00}, so the line plotted is a guide to the slope
	only.
	}
\label{fig_BB00_scaling}
\end{figure}

\begin{figure}[htb]
\plotone{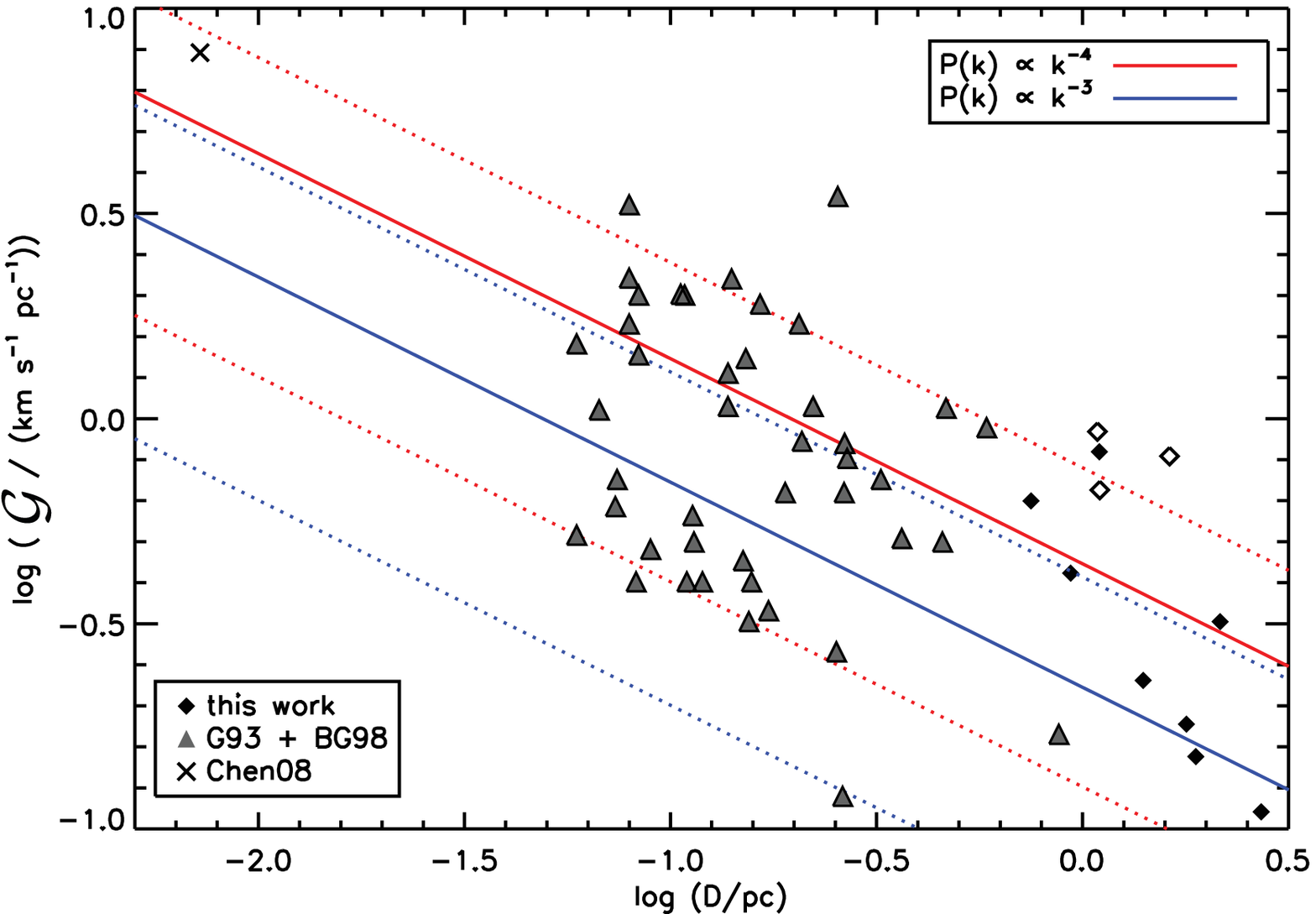}
\caption{The gradient measured for each extinction region versus the diameter
	of the region (diamonds).
	The open diamonds show the results for the three
	extinction regions where the \thirco coverage was less than 80\%
	and hence less reliable.  The triangles show the data from the
	NH$_3$ cores of \citet{Goodman93}, while the cross shows
	data from \citet{Chen08}; see text for details.
	The red diagonal
	lines show the range (dotted) and most likely values 
	(solid)
	predicted by BB00 for a $k^{-4}$ turbulent
	power spectrum, while the blue lines show the range
	(dotted) and most likely values 
	(solid) for a $k^{-3}$ turbulent
	power spectrum.  
}
\label{fig_grad_vs_BB00}
\end{figure}

Although the BB00 models were run under the assumption of a single
size scale (cores with a radius of 0.1~pc), 
BB00 assume a Larson-type scaling law with 
$\sigma \propto R^{0.5}$ to 
compare their results to
that of the entire \citetalias{Goodman93} sample of NH$_3$ cores.
Figure~\ref{fig_BB00_scaling} shows the velocity
dispersions and diameters of the extinction regions
we measure as well as the G93 NH$_3$ cores. 
The extinction regions follow a similar trend to the
G93 cores, hence the BB00 models can
be compared to the extinction regions in the same
manner as was done for the G93 cores.
In Figure~\ref{fig_grad_vs_BB00}, we show the BB00 
predictions for gradients resulting from turbulent
power spectra with $P(k) \propto k^{-4}$ and $P(k) \propto k^{-3}$
(diagonal lines), as well as the \citetalias{Goodman93} data (triangles)
and our own \thirco measurements (diamonds).
For comparison, we also show a recent measurement of 
the size and velocity gradient of an embedded protostar (IRS 1)
within the isolated Bok globule, BHR~71 \citep{Chen08}.
To plot the \citetalias{Goodman93}
data, we adopt the same procedure as BB00 -- the diameter used 
is the geometric mean of the \citetalias{Goodman93} FWHM
sizes of the major and minor axes, and updated gradient measurements
for the seven sources in \citetalias{Barranco98} replace those from
\citetalias{Goodman93}.  The
solid diagonal lines indicate the most likely values of the
gradient found in the BB00 simulations, while the dotted lines
indicate the half maximum likelihood values.  While the extinction regions
we analyze are larger than the cores in \citetalias{Goodman93}, our
observations clearly follow the same trend predicted by the
BB00 model.  Although the centrally concentrated density distribution
adopted in the BB00 model is not a good representation of the
density distribution in the 
regions we analyze, BB00 state that their results are not
strongly dependent on the density distribution -- they found 
similar results when a constant density distribution was adopted.
The extension of their predictions to the \thirco gradients in
the extinction regions measured here is therefore reasonable.
{\hknew A generalized formulation of the velocity gradient
for various density profiles and turbulent power spectra is given
in \citet{Kratter06}, and also shows similar results are
expected regardless of the assumed density distribution.}

\section{DISCUSSION -- ENERGETICS}
The above analyses all point towards a common picture of the
energetics of dense cores within the context of the molecular
clouds they inhabit.  
Previous work \citepalias[e.g.,][]{Kirk07} has shown that dense
cores have little motion with respect to their immediate
surrounding gas.  Our analysis comparing the \nh and \thirco
data has shown that the small motions
continue to larger scales.
Within the larger extinction
regions that the cores inhabit, the cores have 
smaller core-to-core velocity dispersions than the typical
\thirco gas.
Interpretation of this observation is hampered by the lack of knowledge of
the full 3D structure of the lower density cloud material. 
The lower density gas traced by \thirco may be tracing 
multiple structures (or a larger length scale) 
along the line of sight that the cores are not
associated with.

The large-scale low density gas appears to 
be dominated by turbulent motions.  
Across each extinction region, the gradients in \thirco
centroid velocity were measured and
appear to account for roughly half of the total velocity
dispersion of the region.  
This is approximately the same size as is
predicted to arise from large-scale modes of a turbulent
power spectrum with $P(k) \propto k^{-4}$ (BB00).
The velocities of the dense cores within each
extinction region follow the velocity gradient found in
the \thirco gas.  This is consistent with a picture where
both the dense cores and \thirco gas obey Larson's Laws; the
cores trace a smaller volume of gas than the \thirco both
across the plane of the sky and along the line of sight,
and so show a smaller core-to-core velocity dispersion 
than the regional \thirco velocity dispersion, while still
tracing the same large-scale velocity gradient.
A direct comparison of the ratio in the core-to-core and
extinction region velocity dispersions with the prediction from Larson's
Laws could be made once similar 
measurements exist for several molecular clouds.  
Determining this effect quantitatively for a single
region is complicated by the lack of knowledge of the line
of sight distribution of cores and \thircons.  With a large
enough sample of regions, however, this effect should be 
quantifiable across the ensemble.

Comparison of the data to expectations from virial
equilibrium is unclear at best.  As usual, the surface terms
of the virial equation are unknown and must be excluded.  In addition, 
the gas within each extinction region must be assumed to be a single
coherent entity both across the sky and along the line of
sight and the substantial velocity gradient across
each region discussed above must also be ignored.  
With these many assumptions, the virial velocity
dispersion of each region could then be estimated in some of the
traditional ways -- using the mass and size of each region
(as in KTJ07) or using the \thirco velocity dispersion as a 
proxy (similar to Larson 1981).  The core-to-core motion
within each region would then be interpreted as ``sub-virial'',
since their velocity dispersion is much smaller than either
of these two estimates of the region's virial velocity dispersion.
Given the questionable assumptions that are required to
make the virial argument, in particular ignoring the large-scale
gradient across each region and the possibility of multiple
structures along the line of sight, we do not recommend interpreting
the data under this framework.  Instead, future `observations' of numerical
simulations will be able to reveal the range of physical conditions
that can match our results.



\section{CONCLUSIONS}
We analyze the kinematic relationship between dense cores and the
larger environments they inhabit using pointed \nh and \co observations
\citepalias{Kirk07} as well as a \thirco spectral cube 
\citep{Ridge06,Pineda08}
of the Perseus molecular cloud.  
We present a series of kinematic measures complementing those
in \citepalias{Kirk07} that can be used to test models of star
formation.  {\hknew  We choose to investigate physically-motivated,
but procedurally straightforward measures of kinematics in Perseus.
Through this, we hope that future simulations and as well as observations
of other star-forming regions will adopt similar measures to
allow constraints of models of star formation across differing
environments.}

We find the dense cores
have small motions with respect to the \thirco gas 
along the same line of sight,
with a standard deviation of 0.33~km~s$^{-1}$, 
or 0.39 times the \thirco velocity dispersion at
that location, while
the sound speed in the ambient medium is $\sim 0.23$~km~s$^{-1}$.
The small motions we measure are 
an upper limit on the true motion between the \nh and \thircons, 
since the \thirco data
have a beamsize roughly twice the size of the \nh data.  
{\hknew The \thirco may also be tracing additional kinematic
structures that are not traced by \nhns.}
The core-to-core motions within the larger extinction regions
they inhabit are about half of the total (\thircons) velocity dispersion
within the region.
The dense cores tend to follow the overall velocity
gradient across each extinction region to the same
extent as the \thirco gas.  The gradients across
each extinction region tend to account for roughly
half of the total measured velocity dispersion,
and are of the magnitude as expected by large-scale
turbulent modes following a power spectrum of
$P(k) \propto k^{-4}$ \citep{Burkert00}.
The kinematic measures presented here should be relatively
straighforward to obtain both in other large observational datasets
as well as numerical simulations of star formation.  Together
with the measures presented in \citetalias{Kirk07}, we provide
a set of benchmarks by which future star formation models can
be evaluated.

\acknowledgements
We would like to thank Stella Offner for partipating in
a series of lively discussions amongst the authors on various projection
effects and for providing a quick analysis of a core in a 3D simulation
in order to test several ideas.  HK also thanks Fabian Heitsch for
an informative discussion on velocity gradients caused by large-scale
modes of turbulence.  Finally, we would like to thank the referee
for a thorough and thoughtful report which improved our paper,
particularly in the statistics presented.

HK is supported by a Natural Sciences and Engineering Research Council
of Canada Postdoctoral Fellowship, with additional support from the
SAO.  DJ is supported by a Natural
Sciences and Engineering Research Council of Canada grant.
JEP is supported by the NSF through grant \#AF002 from the 
Association of Universities for Research in Astronomy, Inc., 
under NSF cooperative agreement AST-9613615 and by Fundaci\'on Andes 
under project No. C-13442.  Support for this work was provided by the 
NSF through awards GSSP06-0015 and GSSP08-0031 from the NRAO.
This material is based upon work supported by the National Science 
Foundation under Grant No. AST-0407172 and AST-0908159 to AAG.

\appendix
\section{GLOSSARY OF KINEMATIC MEASURES}
A large number of kinematic measures are presented in this paper
in order to provide a variety of diagnostics on the molecular cloud.
For clarity, Table~\ref{tab_defs} summarizes all of these kinematic
measures, along with a brief description of their meaning.

\clearpage
\input{tab1}
\input{tab2}
\input{tab3}
\input{tab4}
\input{tab5}

\clearpage


\clearpage

\end{document}

%% file: tab1.tex
\begin{deluxetable}{cccccccccc}
\tabletypesize{\scriptsize}
\tablewidth{0pt}
\tablecolumns{10}
\tablecaption{\nh to CO Relative Motions \label{tab_rel_motions} }
\tablehead{
\colhead{} &
\multicolumn{2}{c}{\co \tablenotemark{a}} &
\multicolumn{3}{c}{\thirco point \tablenotemark{b}} &
\colhead{} &
\multicolumn{3}{c}{\thirco region \tablenotemark{d}} 
\\
\colhead{\nh\tablenotemark{a}} &
\colhead{$v_{N2H+} - v_{C18O}$ } &
\colhead{$\sigma_{C18O}$ } &
\colhead{$v_{N2H+} - v_{13CO}$ } &
\colhead{$\sigma_{13CO}$} &
\colhead{Good} &
\colhead{Region\tablenotemark{c}} &
\colhead{$v_{N2H+} - v_{13CO,reg}$} &
\colhead{$\sigma_{13CO,reg}$} &
\colhead{Good} 
\\
\colhead{\#} &
\colhead{(km~s$^{-1}$)} & 
\colhead{(km~s$^{-1}$)} & 
\colhead{(km~s$^{-1}$)} & 
\colhead{(km~s$^{-1}$)} & 
\colhead{Fit?} & 
\colhead{\#} & 
\colhead{(km~s$^{-1}$)} & 
\colhead{(km~s$^{-1}$)} & 
\colhead{Fit?} 
}
\startdata
   4 &  0.13 &  0.91 &  0.11 &  0.58 &    Y & 1 &  0.19 &  0.64 &    Y \\
   5 &  0.01 &  1.00 &  0.21 &  0.64 &    Y & 1 &  0.15 &  0.64 &    Y \\
   6 &  0.19 &  0.86 &  0.35 &  0.67 &    Y & 1 &  0.32 &  0.64 &    Y \\
   8 & -0.13 &  0.53 & -0.19 &  0.50 &    Y & 1 &  0.02 &  0.64 &    Y \\
  12 & -0.10 &  0.40 &  0.32 &  0.69 &    Y & 1 &  0.71 &  0.64 &    Y \\
  13 &  0.01 &  0.41 &  0.44 &  0.67 &    Y & 1 &  0.94 &  0.64 &    Y \\
  15 & -0.17 &  0.86 & -0.09 &  0.61 &    Y & 2 &  0.14 &  0.91 &    Y \\
  16 &  0.06 &  1.36 &  0.50 &  0.78 &    Y & 2 &  1.07 &  0.91 &    Y \\
  18 &  0.33 &  1.23 &  0.54 &  0.83 &    Y & 2 &  1.12 &  0.91 &    Y \\
  19 & -0.01 &  0.67 &  0.03 &  0.64 &    Y & 2 & -0.25 &  0.91 &    Y \\
  20 &  0.02 &  0.62 &  0.14 &  0.65 &    Y & 2 & -0.23 &  0.91 &    Y \\
  21 &  0.20 &  1.00 &  0.41 &  0.71 &    Y & 2 &  0.22 &  0.91 &    Y \\
  22 & -0.49 &  1.13 & -0.55 &  0.68 &    Y & 2 & -0.69 &  0.91 &    Y \\
  23 &  0.02 &  0.75 & -0.19 &  0.82 &    Y & 2 & -0.53 &  0.91 &    Y \\
  24 & -0.02 &  0.85 &  0.17 &  0.75 &    Y & 2 & -0.06 &  0.91 &    Y \\
  25 &  0.12 &  0.81 &  0.32 &  0.64 &    Y & 2 &  0.26 &  0.91 &    Y \\
  26 &  0.04 &  0.57 & -0.22 &  0.67 &    Y & 2 & -0.24 &  0.91 &    Y \\
  27 & -0.48 &  0.57 & -0.54 &  0.55 &    Y & 2 & -0.56 &  0.91 &    Y \\
  27 & -0.02 &  0.57 & -0.08 &  0.55 &    Y & 2 & -0.10 &  0.91 &    Y \\
  28 & -0.21 &  0.98 & -0.42 &  0.57 &    Y & 2 & -0.34 &  0.91 &    Y \\
  30 &  0.10 &  0.44 &  0.19 &  0.56 &    Y & 2 &  0.04 &  0.91 &    Y \\
  33 &  0.11 &  0.50 &  0.65 &  0.79 &    N & 2 &  0.65 &  0.91 &    Y \\
  34 & -0.03 &  0.58 & -0.20 &  0.63 &    Y & 2 & -0.59 &  0.91 &    Y \\
  36 &  0.03 &  0.35 &  0.26 &  0.61 &    Y & 2 &  0.59 &  0.91 &    Y \\
  41 &  0.06 &  0.39 &  0.39 &  0.89 &    Y & 2 & -0.35 &  0.91 &    Y \\
  53 &  0.10 &  0.50 &  0.36 &  0.65 &    Y & 3 &  0.73 &  1.01 &    Y \\
  55 &  0.06 &  0.50 &  0.57 &  0.90 &    N & 3 &  0.45 &  1.01 &    Y \\
  63 & -0.06 &  0.64 & -0.08 &  0.70 &    Y & 4 & -0.35 &  0.88 &    Y \\
  67 & -0.40 &  1.00 & -0.66 &  0.71 &    Y & 5 & -0.46 &  0.89 &    Y \\
  68 &  0.11 &  1.04 &  0.31 &  0.92 &    Y & 5 &  0.23 &  0.89 &    Y \\
  71 &  0.25 &  0.85 & -0.07 &  0.85 &    Y & 5 & -0.03 &  0.89 &    Y \\
  72 & -0.20 &  1.20 & -0.38 &  0.94 &    Y & 5 & -0.42 &  0.89 &    Y \\
  73 & -0.22 &  0.95 & -0.24 &  0.81 &    Y & 5 & -0.26 &  0.89 &    Y \\
  74 & -0.09 &  0.86 & -0.18 &  0.70 &    Y & 5 &  0.15 &  0.89 &    Y \\
  75 & -0.07 &  0.89 & -0.16 &  0.83 &    Y & 5 & -0.06 &  0.89 &    Y \\
  76 & -0.12 &  0.73 & -0.20 &  0.61 &    Y & 5 & -0.18 &  0.89 &    Y \\
  77 & -0.04 &  0.79 & -0.15 &  0.84 &    Y & 5 & -0.05 &  0.89 &    Y \\
  78 & -0.01 &  1.09 & -0.19 &  0.91 &    Y & 5 & -0.02 &  0.89 &    Y \\
  79 &  0.25 &  1.02 &  0.09 &  0.91 &    Y & 5 &  0.13 &  0.89 &    Y \\
  84 & -0.21 &  0.79 &  0.23 &  0.96 &    Y & 5 & -0.08 &  0.89 &    Y \\
  85 & -0.09 &  0.98 & -0.16 &  0.71 &    Y & 5 & -0.28 &  0.89 &    Y \\
  86 &  0.01 &  0.96 &  0.25 &  0.74 &    Y & 5 &  0.21 &  0.89 &    Y \\
  87 &  0.00 &  0.41 & -0.48 &  0.85 &    Y & 5 & -0.79 &  0.89 &    Y \\
  90 & -0.07 &  1.08 & -0.19 &  0.94 &    Y & 6 &  0.41 &  0.91 &    Y \\
  91 &  0.05 &  0.56 &  0.68 &  0.87 &    Y & -- & -- & -- &    N \\
  92 &  0.02 &  0.57 &    -- &    -- &   -- & 6 & -0.44 &  0.91 &    Y \\
  93 & -0.23 &  0.60 &    -- &    -- &   -- & 6 & -0.58 &  0.91 &    Y \\
  94 & -0.08 &  0.67 &    -- &    -- &   -- & 6 & -0.62 &  0.91 &    Y \\
  95 &  0.05 &  0.52 & -0.07 &  0.39 &    Y & 7 &  0.53 &  1.25 &    Y \\
  96 &  0.01 &  0.71 & -0.10 &  1.17 &    N & 7 & -0.08 &  1.25 &    Y \\
  97 & -0.01 &  0.49 & -0.00 &  0.48 &    Y & 7 & -0.12 &  1.25 &    Y \\
  98 &  0.08 &  1.05 & -0.29 &  0.91 &    Y & 7 & -0.15 &  1.25 &    Y \\
  99 &  0.01 &  0.39 & -0.09 &  0.74 &    N & 7 & -0.08 &  1.25 &    Y \\
  99 &  0.00 &  0.46 & -0.72 &  0.74 &    N & 7 & -0.71 &  1.25 &    Y \\
 100 & -0.03 &  0.71 & -0.14 &  0.81 &    Y & 7 & -0.10 &  1.25 &    Y \\
 101 & -0.14 &  1.42 &  0.27 &  1.15 &    Y & 7 &  0.59 &  1.25 &    Y \\
 102 &  0.12 &  1.11 &  0.18 &  1.04 &    N & 7 &  0.22 &  1.25 &    Y \\
 103 &  0.19 &  0.86 & -0.58 &  1.22 &    Y & 7 & -0.83 &  1.25 &    Y \\
 103 &  0.55 &  0.86 & -0.22 &  1.22 &    Y & 7 & -0.47 &  1.25 &    Y \\
 104 &  0.05 &  1.41 &  0.88 &  1.57 &    Y & 7 &  0.97 &  1.25 &    Y \\
 106 &  0.19 &  1.92 &  0.06 &  1.19 &    Y & 7 & -0.13 &  1.25 &    Y \\
 107 &  0.25 &  0.41 & -0.38 &  1.22 &    Y & 7 & -0.38 &  1.25 &    Y \\
 109 & -0.04 &  1.49 &  0.62 &  1.51 &    Y & 7 &  0.86 &  1.25 &    Y \\
 110 &  0.41 &  1.88 &  0.43 &  1.26 &    Y & 7 &  0.35 &  1.25 &    Y \\
 111 &  0.14 &  1.59 & -0.52 &  1.25 &    Y & 7 & -0.80 &  1.25 &    Y \\
 111 &  0.35 &  2.13 &  0.40 &  1.25 &    Y & 7 &  0.12 &  1.25 &    Y \\
 112 &  0.22 &  1.51 &  0.65 &  1.38 &    Y & 7 &  0.78 &  1.25 &    Y \\
 113 &  0.24 &  1.46 &  0.36 &  1.18 &    Y & 7 &  0.35 &  1.25 &    Y \\
 115 & -0.30 &  1.53 & -0.20 &  1.22 &    Y & 7 & -0.46 &  1.25 &    Y \\
 116 & -0.17 &  1.53 &  0.04 &  1.01 &    Y & 7 &  0.21 &  1.25 &    Y \\
 118 & -0.28 &  1.31 & -0.27 &  1.21 &    Y & 7 & -0.28 &  1.25 &    Y \\
 118 &  0.12 &  1.31 &  0.13 &  1.21 &    Y & 7 &  0.12 &  1.25 &    Y \\
 121 & -0.06 &  0.92 &  0.20 &  0.94 &    Y & 7 & -0.40 &  1.25 &    Y \\
 122 & -0.08 &  0.96 & -0.29 &  0.74 &    N & 7 &  0.36 &  1.25 &    Y \\
 123 & -0.15 &  0.94 & -0.18 &  0.72 &    N & 7 &  0.57 &  1.25 &    Y \\
 124 &  0.13 &  1.57 &  0.06 &  0.98 &    Y & 7 & -0.58 &  1.25 &    Y \\
 125 & -0.07 &  0.70 & -0.46 &  0.93 &    Y & 7 & -0.34 &  1.25 &    Y \\
 126 & -0.02 &  1.64 & -0.17 &  1.12 &    Y & 7 & -0.78 &  1.25 &    Y \\
 127 &  0.04 &  0.43 & -0.02 &  0.46 &    Y & 9 &  0.18 &  0.87 &    Y \\
 128 &  0.07 &  1.40 & -0.28 &  1.13 &    N & 7 & -0.41 &  1.25 &    Y \\
 129 &  0.01 &  0.23 &    -- &    -- &   -- & 9 & -0.80 &  0.87 &    Y \\
 130 & -0.01 &  0.67 &    -- &    -- &   -- & 9 & -0.51 &  0.87 &    Y \\
 131 &  0.05 &  0.38 &    -- &    -- &   -- & 9 & -0.58 &  0.87 &    Y \\
 132 & -0.37 &  1.01 &    -- &    -- &   -- & 9 & -0.59 &  0.87 &    Y \\
 133 & -0.16 &  1.07 &    -- &    -- &   -- & 9 & -0.30 &  0.87 &    Y \\
 134 & -0.04 &  0.70 &    -- &    -- &   -- & 9 & -0.49 &  0.87 &    Y \\
 135 &  0.04 &  0.70 &    -- &    -- &   -- & 9 & -0.64 &  0.87 &    Y \\
 136 &  0.24 &  1.34 &    -- &    -- &   -- & 9 & -0.72 &  0.87 &    Y \\
 136 &  0.04 &  0.45 &    -- &    -- &   -- & 9 &  0.58 &  0.87 &    Y \\
 138 & -0.17 &  0.90 &    -- &    -- &   -- & 9 & -0.29 &  0.87 &    Y \\
 139 &  0.17 &  0.97 &    -- &    -- &   -- & 9 & -0.27 &  0.87 &    Y \\
 140 &  0.00 &  0.27 &    -- &    -- &   -- & 9 & -0.02 &  0.87 &    Y \\
 143 &  0.00 &  0.46 &  0.12 &  0.56 &    Y & 9 & -0.27 &  0.87 &    Y \\
 145 & -0.18 &  1.31 & -0.42 &  0.60 &    N &10 & -0.94 &  0.81 &    Y \\
 146 &  0.15 &  0.88 &  0.06 &  0.71 &    Y &10 &  0.13 &  0.81 &    Y \\
 147 &  0.09 &  0.86 &  0.05 &  0.70 &    Y &10 &  0.28 &  0.81 &    Y \\
 148 & -0.25 &  0.53 &  0.35 &  0.80 &    Y &10 &  0.75 &  0.81 &    Y \\
 148 &  0.22 &  1.02 &  0.27 &  0.80 &    Y &10 &  0.67 &  0.81 &    Y \\
 149 &  0.16 &  1.07 &  0.01 &  0.79 &    N &10 &  0.15 &  0.81 &    Y \\
 150 & -0.04 &  0.68 & -0.20 &  0.74 &    N &10 & -0.31 &  0.81 &    Y \\
 151 &  0.11 &  0.56 &  0.08 &  0.47 &    Y &11 &  0.23 &  0.88 &    Y \\
 152 &  0.03 &  0.74 & -0.04 &  0.62 &    N &10 & -0.28 &  0.81 &    Y \\
 155 &  0.04 &  0.49 & -0.20 &  0.45 &    N &11 & -0.14 &  0.88 &    Y \\
\enddata
\tablenotetext{a}{\nh core number, \nh to \co relative velocity, and \co velocity dispersion from \citet{Kirk07}.}
\tablenotetext{b}{\nh to \thirco relative velocity, \thirco velocity dispersion at the point of the \nh observation, and whether the \thirco fit was deemed good.  Only good fits were used in the analysis.Dashes denote regions where \thirco data does not exist.}
\tablenotetext{c}{Associated extinction region.}
\tablenotetext{d}{\nh to \thirco relative velocity, \thirco linewidth for  the region and whether the \thirco fit was deemed good.  Only good fits were used in the analysis.}
\end{deluxetable}

%% file: tab2.tex
\begin{deluxetable}{ccccc}
\tablecolumns{5}
\tablecaption{Region Velocity Dispersions in \thirco \label{tab_reg_widths} }
\tablehead{
\colhead{Region \tablenotemark{a} } &
\colhead{Areal \tablenotemark{a} } &
\colhead{$\sigma_{reg,Gauss}$ \tablenotemark{b} }  &
\colhead{Good \tablenotemark{b} } &
\colhead{$\sigma_{reg,FWQM}$ \tablenotemark{c} } 
\\
\colhead{\#} & 
\colhead{Cov. (\%)} &
\colhead{(km~s$^{-1}$)} &
\colhead{Fit?} & 
\colhead{(km~s$^{-1}$)} 
}
\startdata
  1 & 100 &  0.64 &  Y &  0.60  \\
  2 & 100 &  0.91 &  Y &  0.88  \\
  3 &  95 &  1.01 &  N &  0.98  \\
  4 &  55 &  0.88 &  N &  1.06  \\
  5 & 100 &  0.89 &  Y &  0.86  \\
  6 &  52 &  0.91 &  Y &  0.86  \\
  7 & 100 &  1.25 &  N &  1.22  \\
  8 & 100 &  0.90 &  Y &  0.84  \\
  9 &  40 &  0.87 &  Y &  0.82  \\
 10 & 100 &  0.81 &  Y &  0.74  \\
 11 &  84 &  0.88 &  Y &  0.84  \\
\enddata
\tablenotetext{a}{Extinction region number and fractional coverage of the area observed in \thirco over the extinction region.}
\tablenotetext{b}{The Gaussian-fit \thirco region velocity dispersion and whether the fit was judged to be good.}
\tablenotetext{c}{The equivalent \thirco region velocity dispersion as measured from the FWQM.  See text for details.}
\end{deluxetable}

%% file: tab3.tex
\begin{deluxetable}{cccc}
\tablewidth{0pt}
\tablecolumns{4}
\tablecaption{Core-to-Core Velocity Dispersions \label{tab_core_core} }
\tablehead{
\colhead{Region} &
\colhead{$\sigma v_{sl,A}$ \tablenotemark{a}} &
\colhead{$\sigma v_{all,A}$ \tablenotemark{a}} &
\colhead{$\sigma v_{all,B}$ \tablenotemark{b}} 
\\
\colhead{\#} &
\colhead{(km~s$^{-1}$)} &
\colhead{(km~s$^{-1}$)} &
\colhead{(km~s$^{-1}$)} 
}
\startdata
  1 &  0.39  &  0.36  &  0.35  \\
  2 &  0.55  &  0.54  &  0.31  \\
  3 &  0.20  &  0.20  &  0.20  \\
  4 &    --  &    --  &  0.14  \\
  5 &  0.28  &  0.28  &  0.21  \\
  6 &  0.09  &  0.49  &  0.20  \\
  7 &  0.54  &  0.50  &  0.56  \\
  8 &    --  &    --  &    --  \\
  9 &  0.32  &  0.38  &  0.41  \\
 10 &  0.55  &  0.56  &  0.35  \\
 11 &  0.26  &  0.26  &  0.26  \\
\enddata
\tablenotetext{a}{Core-to-core velocity dispersion derived for starless cores and all cores using method A. Dashes denote regions with less than two cores.  See text for details.}
\tablenotetext{b}{Core-to-core velocity dispersion derived for all cores using method B.  Dashes denote regions with no cores.  See text for details.}
\end{deluxetable}

%% file: tab4.tex
\begin{deluxetable}{ccccccc}
\tablecolumns{7}
\tablecaption{\thirco Gradients Across Each Extinction Region \label{tab_tot_grads} }
\tablehead{
\colhead{Region \tablenotemark{a}} &
\colhead{$\mathcal{G}$ \tablenotemark{b}} &
\colhead{Error \tablenotemark{b}} &
\colhead{Angle \tablenotemark{c}} &
\colhead{Error \tablenotemark{c}} &
\colhead{$D_{reg}$ \tablenotemark{d}} &
\colhead{$\mid \mathcal{G} \mid D_{reg}$ \tablenotemark{e}} \\ 
\colhead{ \#} &
\colhead{(km~s~$^{-1}$~pc~$^{-1}$)} &
\colhead{(km~s~$^{-1}$~pc~$^{-1}$)} &
\colhead{(degrees)} & 
\colhead{(degrees)} & 
\colhead{(pc)} &
\colhead{(km~s$^{-1}$)}
}
\startdata
  1 &0.1456 & 0.0009 &   84.3 & 0.3 & 1.88 & 0.27 \\
  2 &0.1115 & 0.0002 & -118.78 & 0.08 & 2.71 & 0.30 \\
  3 &0.1845 & 0.0007 &   84.5  & 0.2 & 1.79 & 0.33 \\
  4 &0.811  & 0.002  & -132.5 & 0.1 & 1.63 & 1.32 \\
  5 &0.2252 & 0.0004 &  -82.98 & 0.07 & 1.40 & 0.32 \\
  6 &0.667  & 0.005  &   23.4 & 0.3 & 1.10 & 0.73 \\
  7 &0.3230 & 0.0003 &   -6.11 & 0.04 & 2.16 & 0.70 \\
  8 &0.827  & 0.003  & -48.6 & 0.1 & 1.10 & 0.91 \\
  9 &0.934  & 0.005  & -56.1 & 0.2 & 1.09 & 1.01 \\
 10 &0.420  & 0.002  & -42.3 & 0.2 & 0.94 & 0.39 \\
 11 &0.627  & 0.002  &  -33.4 & 0.1 & 0.75 & 0.47 \\
\enddata
\tablenotetext{a}{The extinction region number}
\tablenotetext{b}{The velocity gradient and associated error measured across each extinction region.  See Section~6 for more detail.}
\tablenotetext{c}{Orientation of the gradient and associated error.  The angle is given in degrees clockwise from North.}
\tablenotetext{d}{Extinction region diameter from \citetalias{Kirk06}.}
\tablenotetext{e}{The velocity dispersion inferred from the gradient across each region.}
\end{deluxetable}

%% file: tab5.tex
\begin{deluxetable}{ccc}
\rotate
\tabletypesize{\small}
\tablewidth{0pt}
\tablecolumns{3}
\tablecaption{Summary of Kinematic Measures Used \label{tab_defs} }
\tablehead{
\colhead{Symbol} &
\colhead{Formula} &
\colhead{Description}
}
\startdata
$v_x$	& & Centroid velocity of species x \\
$\sigma_{x}$ &  & Velocity dispersion (Gaussian sigma) of species x \\
   & $v_{N2H+} - v_{x}$ & Relative motion between \nh and species x \\
$\zeta_{norm,x}$ & $\frac{v_{N2H+} - v_x}{\sigma_x}$ & Normalized 
relative motion\\
$\sigma_{reg, Gauss}$ & & Region velocity dispersion using a Gaussian fit
\tablenotemark{a} \\
$\sigma_{reg, FWQM}$ & $ FWQM_{reg} / 4 ln \sqrt{2} $ & Region velocity 
dispersion
measured using the full width at quarter maximum \tablenotemark{a} \\
$\sigma v_{meth. A}$ & $\sigma(v) $ & 
Dispersion of core centroid velocities within an extinction region 
(``method A'') \\
$\sigma v_{meth. B}$ & & Velocity dispersion of the cores within an extinction 
region (``method B'') \tablenotemark{b} \\
$D_{reg}$ & & Diameter of extinction region \\
$ \mathcal{G} $ & & Velocity gradient across a region \\
$(x_c,y_c)$ &  & position of centre of region used in gradient calculation \\
$d_{\parallel}$ & & Distance between $(x,y)$ and $(x_c,y_c)$ parallel to the
	gradient \\
$v_{flow,mod}(x,y)$ & $v(x_c,y_c) + \mid {\mathbf {\mathcal G}} \mid d_{\parallel} $ &
Velocity inferred from the bulk gradient \\
$\sigma_{reg,mod}$ & $\mathcal{G} \times D_{reg}$ & Region velocity dispersion
	inferred from the bulk gradient \\
\enddata
\tablenotetext{a}{Measured using the cumulative spectrum of the entire 
	extinction region}
\tablenotetext{b}{Measured using the cumulative core spectra within the
	extinction region}

\end{deluxetable}